\documentclass[lettersize,journal]{IEEEtran}
\usepackage{amsmath,amsfonts}
\usepackage{algorithm}
\usepackage{algpseudocode}
\usepackage{array}
\usepackage{textcomp}
\usepackage{url}
\usepackage{verbatim}
\usepackage{graphicx}
\usepackage{cite}
\usepackage{caption}
\usepackage{subcaption}
\usepackage{amssymb}
\usepackage{booktabs}
\usepackage{enumitem}
\usepackage{float}
\usepackage{tabularx}
\usepackage{multirow}
\usepackage{dblfloatfix}
\usepackage[pagebackref,breaklinks]{hyperref}
\usepackage[capitalize]{cleveref}

\newcommand{\mbb}[1]{\mathbb{#1}}
\newcommand{\mcl}[1]{\mathcal{#1}}
\begin{document}
%
\title{High-frequency Matters: An Overwriting Attack and defense for
Image-processing Neural Network Watermarking}
%
%
%
%

\author{Huajie~Chen,
        Tianqing Zhu*,
        Chi Liu,
        Shui Yu,
        and Wanlei Zhou
\thanks{*Tianqing Zhu is the corresponding author.}
\thanks{Huajie Chen, Tianqing Zhu, Chi Liu, and Shui Yu are with the Centre for Cyber Security and Privacy and the School of Computer Science, University of Technology Sydney, Sydney, NSW 2007, Australia (e-mail: huajie.chen@student.uts.edu.au; tianqing.zhu@uts.edu.au; chi.liu@student.uts.edu.au; shui.yu@uts.edu.au;).}
\thanks{Wanlei Zhou is with the Institute of Data Science, City University of Macau, Macao SAR, China (e-mail: wlzhou@cityu.edu.mo).}}

%
%

\markboth{Journal of \LaTeX\ Class Files,~Vol.~14, No.~8, August~2015}%
{Shell \MakeLowercase{\textit{et al.}}: Bare Demo of IEEEtran.cls for Computer Society Journals}
%



\IEEEtitleabstractindextext{%
\begin{abstract}

    In recent years, there has been significant advancement in the field of model watermarking techniques. 
    However, the protection of image-processing neural networks remains a challenge, with only a limited number of methods being developed.
    The objective of these techniques is to embed a watermark in the output images of the target generative network, so that the watermark signal can be detected in the output of a surrogate model obtained through model extraction attacks.
    This promising technique, however, has certain limits. 
    Analysis of the frequency domain reveals that the watermark signal is mainly concealed in the high-frequency components of the output. 
    Thus, we propose an overwriting attack that involves forging another watermark in the output of the generative network.
    The experimental results demonstrate the efficacy of this attack in sabotaging existing watermarking schemes for image-processing networks with an almost 100\% success rate.
    To counter this attack, we propose an adversarial framework for the watermarking network.
    The framework incorporates a specially-designed adversarial training step, where the watermarking network is trained to defend against the overwriting network, thereby enhancing its robustness.
    Additionally, we observe an overfitting phenomenon in the existing watermarking method, which can render it ineffective. 
    To address this issue, we modify the training process to eliminate the overfitting problem.
\end{abstract}

}

\maketitle

\IEEEdisplaynontitleabstractindextext

%
\IEEEpeerreviewmaketitle

\section{Introduction}\label{sec:introduction}

\IEEEPARstart{T}{raining} a high-performing deep learning model is incredibly costly, which consumes a massive amount of computational resources along with electricity, human resources, etc.
In fact, training such models is so time-consuming and expensive, that stealing a model is comparatively much simpler and cheaper.
Beyond directly copying an original model and simply claiming its ownership, there are several genres of deep learning model theft. 
These include model fine-tuning \cite{wang2017growing, nagabandi2018neural, howard2018universal}, model pruning\cite{liu2018rethinking, he2017channel, zhu2017prune}, and knowledge distillation.
In the face of this many potential threats, model owners have sought ways to protect their intellectual properties, and one such method is model watermarking.
Model watermarking is a brand-new technique that embeds a traceable digital watermark into a deep learning model. 
As such, it offers a promising way to model copyright protection.

The first attempt at model watermarking was made in 2017 by Yusuke Uchida {\it et al.} \cite{uchida2017embedding}, who proposed a method of embedding a watermark into a deep learning model.
The watermark was designed to verify ownership of the model given white-box access.
Since then, several exemplary works \cite{rouhani2018deepsigns, szyller2021dawn, namba2019robust} have emerged to provide better protection for deep learning models in different scenarios up to today, where even black-box attacks can be effectively prevented. 

\begin{figure}[t!]
    \centering
    \includegraphics[width=.4\textwidth]{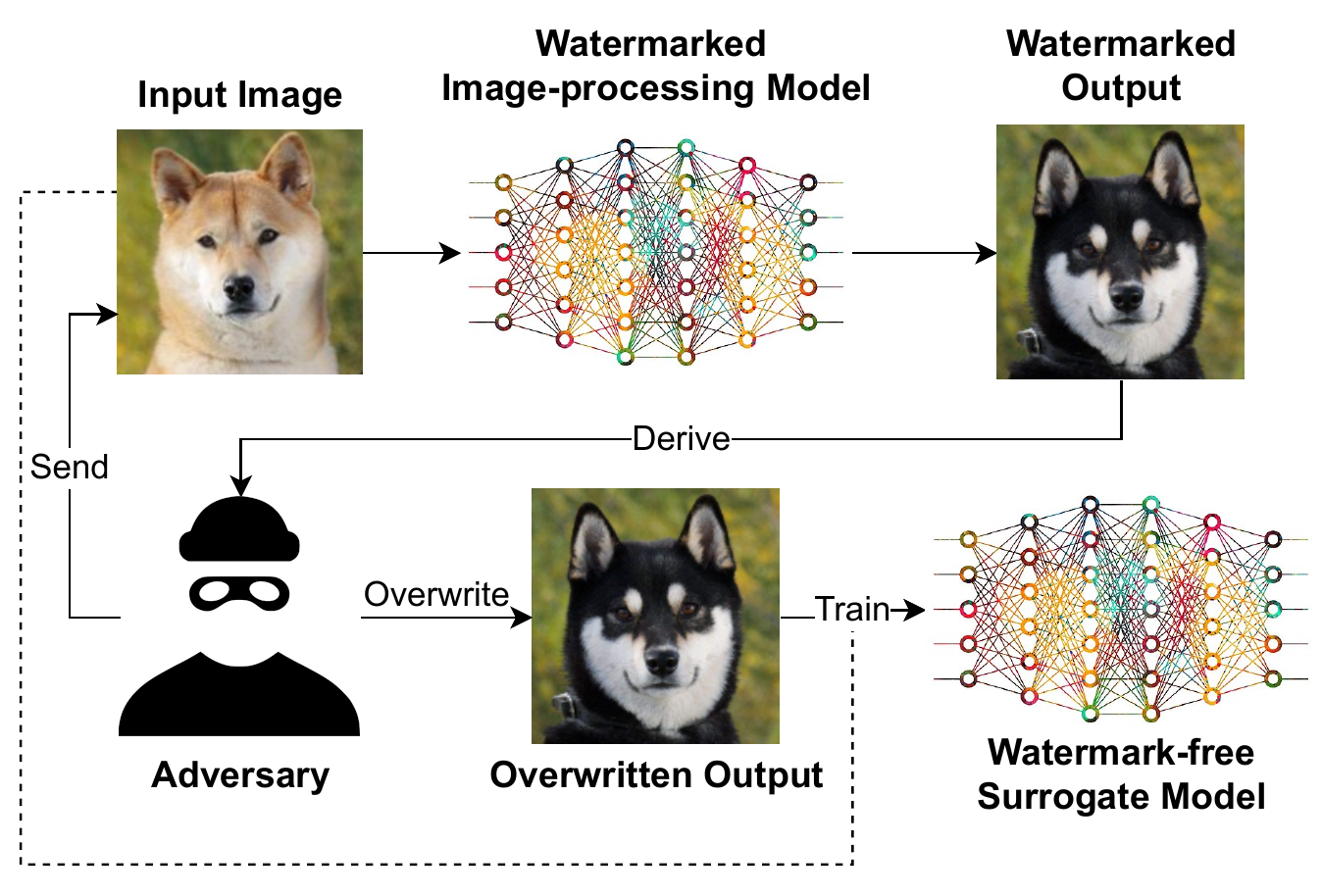}
    \caption{Attack workflow: The adversary sends the input image to the watermarked image-processing model to derive the watermarked output. If it trains a surrogate model directly with the input image and the watermarked output, the surrogate model will contain the watermark information. By overwriting the watermarked output, it removes the watermark in the output set and forges its own watermark inside the overwritten output. Finally, a watermark-free surrogate model can be trained.}
    \label{fig:atk_idea}
\end{figure}

\begin{figure}[t!]
    \centering
    \includegraphics[width=.4\textwidth]{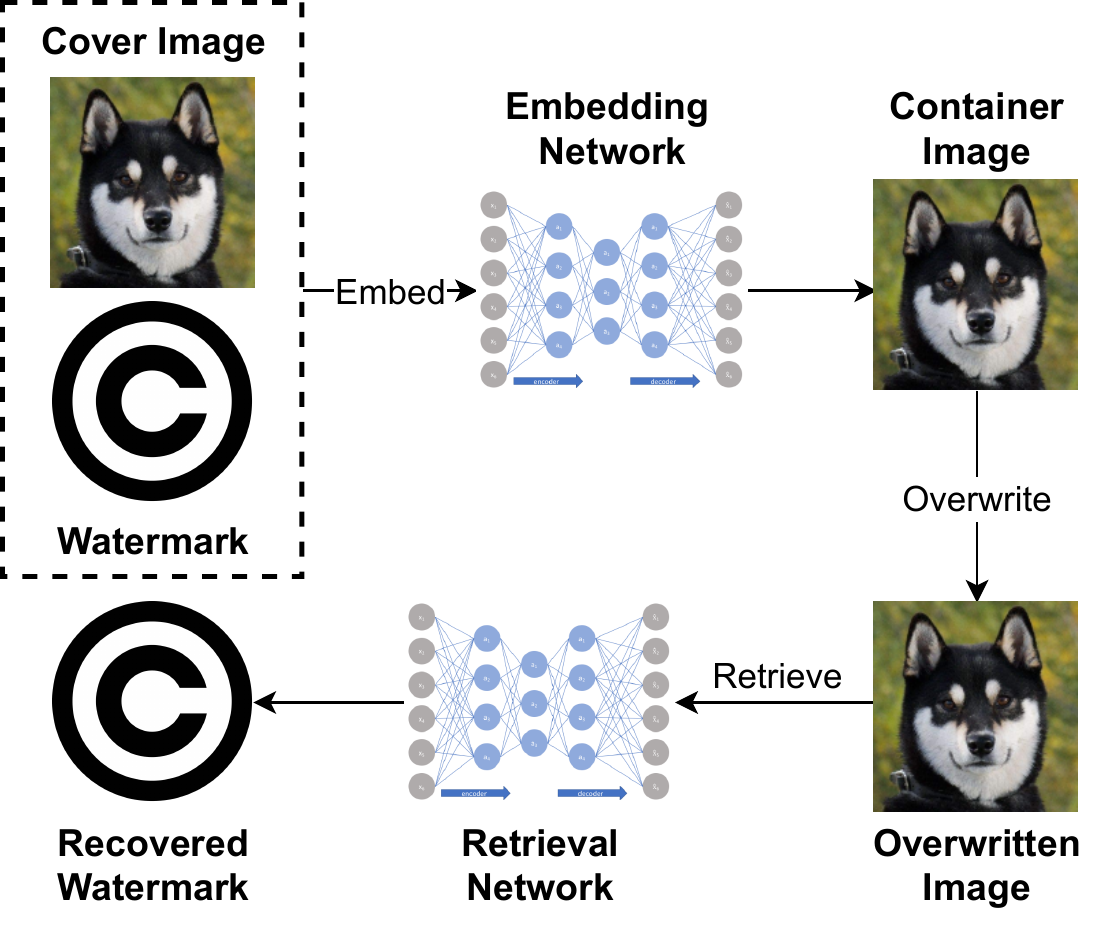}
    \caption{Defense workflow: 
    After embedding the watermark, an overwriting network performs an overwriting attack that yields an overwritten image that is then fed to the retrieval network. 
    The retrieval network is then required to retrieve a valid recovered watermark from the overwritten image. 
    Together, these networks form the defense network.
    }
    \label{fig:def_idea}
\end{figure}

However, most model watermarking approaches are designed to protect classification models; methods that work for image-processing models are few and far between.
In short, an image-processing model takes images as its input and outputs modified images, which is quite unlike a classification model that takes in images and simply outputs the digits of a category.
In 2020, Zhang {\it et al.} \cite{zhang2020model} proposed a framework to watermark image-processing neural networks, which, to the best of our knowledge, was the first work in image-processing model watermarking.
Essentially, Zhang's work combines model watermarking with deep steganography so as to forcibly embed watermark information in the outputs of the released models. 
Deep steganography is a technique that uses deep learning models to hide a secret image completely within a cover image, such that it is invisible to the naked human eye.
The image containing the embedded secret image is called the container image.
By releasing a set of processed images containing a hidden watermark, any attacker intending to steal the model is compelled to train their own watermarked model. 
Subsequently, Quan {\it et al.} \cite{quan2020watermarking} devised another image-processing model watermarking scheme that takes a backdoor watermarking approach.
Briefly, the watermarked model functions normally when it receives normal input images.
When it receives a noise trigger input, it outputs a pre-defined watermark to validate ownership.
Even though recent studies show that steganography plays an essential role in the protection of images, this type of approach might still vulnerable to attacks.



In fact, our study shows that current watermarking methods for image-processing models are not adequately robust. 
For example, we find that, due to the properties of deep steganography, watermarking with image-processing models is vulnerable to changes in the frequency domain, especially the high-frequency domain. 
To outline what we mean, we devise an overwriting attack method that shows how existing image-processing model watermarking methods, and even deep steganography itself can be nullified.
Having designed the attack, we also designed a defense against it that promises to guarantee the safety of deep learning models.
The defense method mitigates the overwriting attack through a new adversarial training framework that combines a watermarking method with the overwriting attack.


The general workflow of the attack is described in Figure \ref{fig:atk_idea}. 
Here, a released image-processing deep learning model is watermarked such that every image it outputs contains an invisible watermark.
If an attacker tries to train a surrogate model via knowledge distillation, the surrogate model will carry the watermark information automatically.
However, in our attack, we train an overwriting network that overwrites the embedded watermark in the output from the watermarked model. 
A surrogate model is also trained with the overwritten output and the input image sets.
Thus, the watermark is nullified, for the original watermark can no longer be retrieved from the output of the surrogate model.


To effectively counter the overwriting attack, we propose an adversarial training framework that deliberately incorporates an overwriting network to enhance the robustness of the watermarking network. 
Figure \ref{fig:def_idea} demonstrates. 
Briefly, an overwriting network is trained along with a watermarking network, which together form a defense network. 
There is an adversarial training process, where the overwriting network tries to overwrite the watermark in the container image so that the retrieval network in the watermarking network cannot retrieve a valid recovered watermark from it. 
In contrast, the watermarking network tries to retrieve a valid recovered watermark even if the container image has been overwritten.
This competitive process significantly boosts the robustness of the watermarking network.

Overall, our contributions are as follows:

\begin{enumerate}[label=\roman*)]
    \item Through frequency analysis, we have unraveled where a secret image signal is embedded in a container image. 
    Accordingly, we devised a possible attack to nullify the currently existing image-processing model watermarking methods. 
    
    \item We devised a corresponding defense method based on adversarial training that counters the proposed attack method with a new adversarial training framework to protect the image-processing network.
    
    
    
    \item We discovered an overfitting problem with the current watermarking method for protecting image-processing models that will nullify the protection, and fixed it by modifying the training process.
\end{enumerate}

The rest of this paper is organized as follows.
In section \ref{sect:pre}, we demonstrate the preliminaries by listing the notations used in the context and illustrating the background and related works.
We then describe our proposed method in detail in Section \ref{sect:method}.
Our experiment processes and results are presented in Section \ref{sect:exp}, and they are analyzed and discussed in Section \ref{sect:dis}.
Lastly, we draw a conclusion about this work in Section \ref{sect:con}.



\section{Preliminary}
\label{sect:pre}



  
    
    
    
    
    

\begin{table}[ht!]
    \caption{Notations}
    \label{tab:notation}
    \begin{tabularx}{.48\textwidth}{
    |>{\centering\arraybackslash}m{.05\textwidth}
    |>{\arraybackslash}m{.377\textwidth}
    |}
    \hline
    
    \multicolumn{1}{|c|}{Notation}
    & 
    \multicolumn{1}{c|}{Definition}
    \\\hline
    
    $\mcl U$
    & 
    The overwriting network.
    \\\hline
    
    $\mcl O$
    & 
    The defense network.
    \\\hline
    
    $\mcl E$
    & 
    An embedding network that embeds a secret image into a cover image to yield a container image.
    \\\hline
    
    $\mcl R$
    & 
    A retrieval network that retrieves a recovered secret image from a container image.
    \\\hline
    
    $\mcl D$
    & 
    A discriminator network that identifies whether or not a given image contains hidden content.
    \\\hline
    
    $\mcl E_{\mcl U}$
    & 
    The overwriting embedding network.
    \\\hline
    
    $\mcl R_{\mcl U}$
    & 
    The overwriting retrieval network.
    \\\hline
    
    $H$
    & 
    The original and watermark-free image-processing model.
    \\\hline
    
    $H'$
    & 
    A surrogate model mimicking $H$, but trained on a watermarked dataset.
    \\\hline
    
    $H_0$
    & 
    A surrogate model mimicking $H$, but trained on a watermark-free dataset.
    \\\hline
    
    $A$
    & 
    A set of images for the image-processing network to process.
    \\\hline
    
    $B$
    & 
    A set of processed images originating from $A$.
    \\\hline
    
    $B'$
    & 
    A set of watermarked and processed images, originating from $B$.
    \\\hline
    
    $B''$
    & 
    A set of noisy output images from, originating from $B$.
    \\\hline
    
    $B_{\mcl U}$
    & 
    A set of watermarked and processed images, but having suffered from the overwriting attack.
    \\\hline
    
    $B_0$
    & 
    A set of processed images from a surrogate model that is not trained on the watermarked dataset.
    \\\hline
    
    $C/c$
    & 
    A set of cover images/a cover image for concealing secrets.
    \\\hline
    
    $C'/c'$
    & 
    A set of container images/a container image where secrets are hidden inside.
    \\\hline
    
    $S/s$
    & 
    A set of secret images/a secret image to hide.
    \\\hline
    
    $S'/s'$
    & 
    A set of recovered secret images/a recovered secret image.
    \\\hline
    
    $w$
    & 
    A watermark.
    \\\hline
    
    $w'$
    & 
    A recovered watermark.
    \\\hline
    
    $w_0$
    & 
    A pure black null image.
    \\\hline
    
    $c'$
    & 
    A container image that contains a watermark.
    \\\hline
    
    $x$
    & 
    An arbitrary image that is the same size as $c'$
    \\\hline
    
    $x'$
    & 
    A recovered image originating from $x$.
    \\\hline
    
    $\epsilon$
    & 
    A tolerable error range of a recovered secret image.
    \\\hline
    
    $\mcl L$
    & 
    A loss function.
    \\\hline
    
    $\lambda$
    & 
    A weight parameter for a regularizer in the loss function.
    \\
    
    \hline
    \end{tabularx}    
\end{table}

\subsection{Watermarking \& Deep Learning}

    


Watermarking is a powerful method for object authentication and ownership validation. It has established strong ties with deep learning in recent times. To provide a comprehensive overview of these interactions, we have categorized them into two main categories: model watermarking and image watermarking using deep learning. For the reader's convenience, a list of all the notations used in the subsequent sections can be found in Table \ref{tab:notation}.

\subsubsection{Model watermarking}

The existing techniques for model watermarking can be classified into three categories: model weight watermarking, backdoor watermarking, and active watermarking.

In model weight watermarking, as described in \cite{uchida2017embedding}, the watermark is embedded into the model's weight parameters during the training process. To retrieve the watermark, one needs complete access to the model's internal structure, which is often not feasible in real-world scenarios. Furthermore, these methods are not highly resilient against attacks such as model pruning, fine-tuning, and knowledge distillation.

Backdoor watermarking, as discussed in \cite{szyller2021dawn}, involves the deliberate alteration of a portion of the training data to create an overfitted model. This portion is referred to as the trigger dataset and can be used to validate the ownership of a suspect model. If the majority of the trigger data result in the suspect model producing the watermark labels, the model's ownership can be confirmed with just black-box access. Compared to model weight watermarking, this method is more robust against the previously mentioned attacks.

On the other hand, active watermarking methods aim to prevent model theft proactively. For instance, Tang et al. \cite{tang2020deep} proposed a method that requires the user to enter a valid serial number before using the desired model. This model is a student model derived from a teacher model and functions correctly only with a valid serial number. Although this approach is proactive in nature and protects the model, a malicious entity can still crack the serial number generator and propagate the stolen model.

\subsubsection{Image watermarking via deep learning}

Image watermarking methods that leverage deep learning can be further categorized into auto-encoder image watermarking and generative adversarial network image watermarking.

Auto-encoder image watermarking, first introduced by Baluja in \cite{baluja2017hiding}, involves the use of an embedding network and a retrieval network.
The embedding network embeds a watermark or secret image into a cover image to produce a container image that is visually similar to the cover image.
The retrieval network then retrieves the watermark from the container image with a tolerable error range.
While these methods achieve high perceptual quality, they are susceptible to steganalysis, a detection attack that identifies hidden content in an image.
Additionally, the container images generated by these methods lack robustness against distortions and malicious attacks that can result in damage or removal of the hidden content.

Generative adversarial network image watermarking is similar to auto-encoder image watermarking, but with the addition of a discriminator in the framework.
During adversarial training, the discriminator is trained to detect hidden content in any image, while the embedding network is tasked with deceiving the discriminator with the container images it generates.
This enhances the covertness of the container images against steganalysis.
However, they remain vulnerable to distortions during transmission and malicious attacks, such as JPEG compression and overwriting attacks.



\subsection{Related Work}

\begin{figure*}[t!]
    \centering
    \includegraphics[width=.9\textwidth]{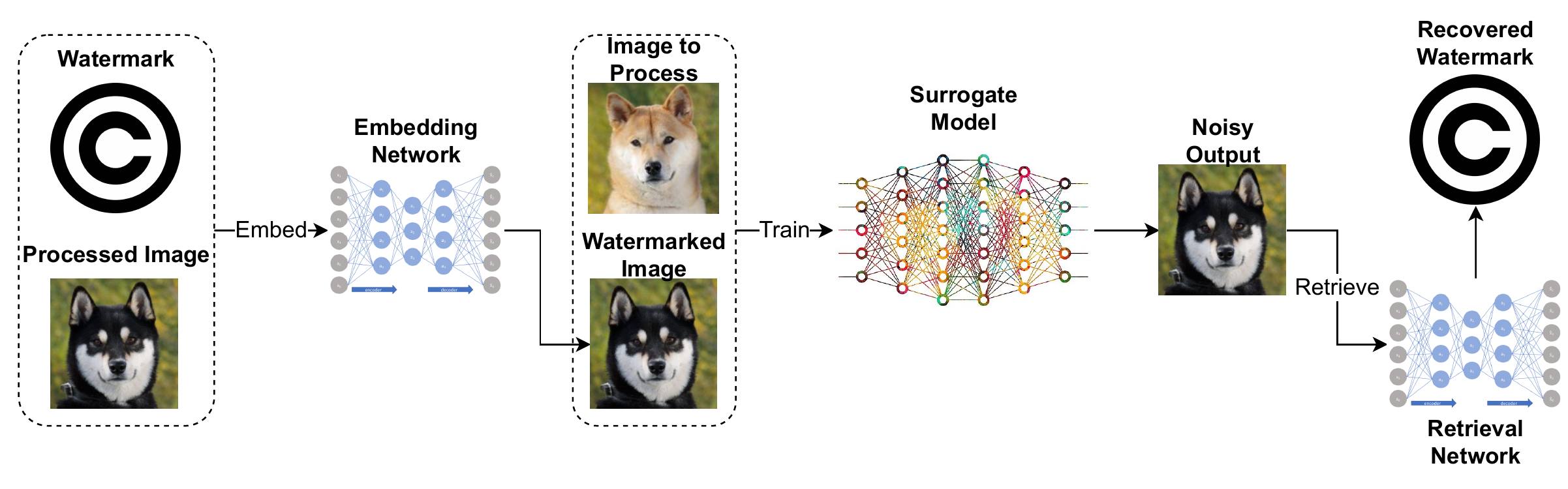}
    \caption{Framework of the Watermarking Network: 
    Starting from the very left, the embedding network is trained to embed a watermark into a processed image set so as to yield a watermarked image set. 
    An adversary trains a surrogate model with a set of raw images and the watermarked image set, and thus the surrogate model carries the watermark information.
    Every time when the surrogate model yields a set of noisy output, the retrieval network is able to retrieve a recovered watermark from the noisy output to validate the model's ownership.
    }
    \label{fig:wm_net}
\end{figure*}


Watermarking is a powerful method for safeguarding intellectual property and preventing copyright infringement in various domains, including image protection \cite{potdar2005survey}, audio \cite{arnold2000audio} and video files \cite{doerr2003guide}. By embedding unique and imperceptible marks within the intellectual property, the watermark serves as evidence of ownership and can be used in legal proceedings to defend against infringement claims. Despite having a long history of use, watermarking is a relatively new application in the realm of deep learning models.

In 2017, Uchida {\it et al.} \cite{uchida2017embedding} introduced a novel method for embedding a watermark into the weight parameters of a model, which was considered to be the first attempt at using watermarking techniques for the protection of intellectual property in neural networks. Despite its pioneering efforts, the method's validity in proving ownership required complete access to the parameters, or white-box access, which made it not practical for real-world scenarios. Furthermore, its robustness to different types of attacks was subject to improvement.

Rouhani {\it et al.} \cite{rouhani2018deepsigns} then proposed a watermarking framework that provides protection against fine-tuning, pruning, and overwriting of watermarks in both white-box and black-box scenarios. This approach was more robust to attacks, however, it was not capable of preventing knowledge distillation attacks.
Szyller {\it et al.} \cite{szyller2021dawn} then introduced an approach that was capable of countering all types of attacks, including knowledge distillation, by making a portion of the output from the watermarked model deliberately false. This strategy forces the surrogate model to include the watermark information by overfitting the falsified labels, thus representing a trade-off between robustness and accuracy.

It is worth noting that all of these methods, including the one proposed by Szyller {\it et al.}, work in a passive manner to defend against attacks, as the watermarks only serve to prove ownership after a copyright violation has already occurred, rather than preventing the violation from happening in the first place. Furthermore, these methods, along with most other model watermarking methods, are designed for classification models, with only a limited number of watermarking methods available for image-processing models.

In 2020, Zhang {\it et al.} proposed a watermarking method for image-processing deep learning models \cite{zhang2020model}. This method is the first of its kind and incorporates the concept of deep steganography, which is the technique of hiding information in such a way that it is not detected. The method fuses imperceptible image watermarking with model watermarking, making it effective against black-box knowledge distillation attacks.
The technique of steganography has a long history, dating back centuries, and has been utilized in different domains. Baluja first introduced the use of a deep learning model for image steganography in 2017 \cite{baluja2017hiding}. The method involves hiding one image within another image in such a way that it is not visible to the naked eye.

Several advancements have been made in the field of deep steganography since then, with Wu {\it et al.} designing a framework to perform end-to-end deep steganography \cite{wu2018image} and Zhang {\it et al.} developing a framework to hide an arbitrary image within another image \cite{zhang2020udh}. In \cite{zhao2022jointw}, an image watermarking method was merged with the image-generative network, using deep steganography to prevent the model from being misused.

However, as deep steganography evolves, so do the attacks. Traditional attacks on steganography include image resizing, cropping, distortion, and compression, as illustrated in Hosam's work \cite{hosam2019attacking}. Additionally, deep learning has been utilized to perform these attacks, as seen in Boroumand {\it et al.}'s work \cite{boroumand2018deep}, where a deep convolution neural network (DCNN) framework was proposed to perform deep steganalysis. Corley {\it et al.} \cite{corley2019destruction} designed a framework based on a generative adversarial network (GAN) with significant performance that is capable of purging secret images hidden in container images. Thus, similar to the battles between attacks and defenses in model watermarking, intense battles also exist in image watermarking through deep steganography.



\section{Method – Attack and defense}
\label{sect:method}

\begin{figure*}[t!]
    \centering
    \includegraphics[width=.7\textwidth]{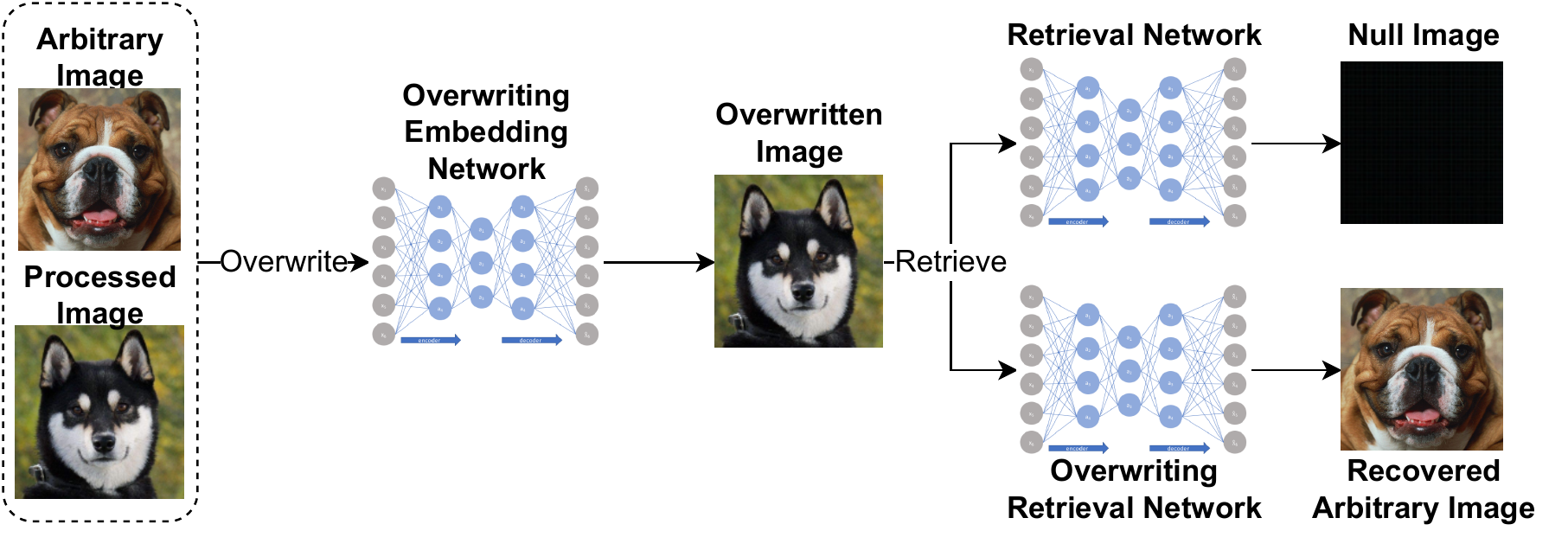}
    \caption{Framework of the Overwriting Network:
    The overwriting network is trained to embed an arbitrary image, or, a watermark, into a container image so as to yield a overwritten image.
    The overwriting also contains a retrieval network that is able to retrieve the recovered image, whereas the retrieval network in the watermarking network can only retrieve a null image from the overwritten image.
    }
    \label{fig:usurper}
\end{figure*}

\subsection{Attack Analysis}
In the current watermarking method for image-processing neural networks, deep steganography is seamlessly integrated with model watermarking.
The watermarking process is composed of two key components, an embedding network $\mathcal{E}$ and a retrieval network $\mathcal{R}$.
As illustrated in Figure \ref{fig:wm_net}, the watermarking process begins by training $\mathcal{E}$ and $\mathcal{R}$ on a set of processed images, $B$, and a watermark image, $w$.
The embedding network $\mathcal{E}$ then embeds the watermark image $w$ into each image $b_i$ in the set $B$ to produce a watermarked image set, $B'$.
This process is denoted as
\begin{equation}
    \begin{aligned}
        B' = \mcl E(B, w).
    \end{aligned}
\end{equation}
In the event of the presence of an adversary, they will only have access to the unprocessed image set $A$, and the watermarked processed image set $B'$. The adversary can then train a surrogate model, denoted as $H'$, using $A$ and $B'$, such that the model learns to produce processed images with watermarks similar to those in $B'$. Finally, the retrieval network $\mcl R$ should be capable of retrieving a recovered watermark $w'$ from both the original watermarked image set $B'$ and the noisy output set $B''$ produced by the surrogate model $H'$, denoted as
\begin{equation}
    \begin{aligned}
        w' = \mcl R(b'), \text{ s.t. } w' = w + \epsilon, \textit{ iff } b' \in B' \cup B'',
    \end{aligned}
\end{equation}
where $\epsilon$ represents a tolerable error range.
Meanwhile, if $\mcl R$ receives a watermark-free image $x$ as input, $\mcl R$ will yield a null image $w_0$ that is purely dark, denoted as 
\begin{equation}
    \begin{aligned}
        w_0 = \mcl R(x), \forall \ x \not \in B' \cup B''.
    \end{aligned}
\end{equation}

However, deep steganography is vulnerable to perturbations in the frequency domain, as highlighted in the work of Zhang {\it et al.} \cite{zhang2021universal}. This motivates us to explore an overwriting attack on the watermarked image set $B'$. The objective of the attack is to generate an overwritten image set $B_{\mcl U}$ such that the retrieval network $\mcl R$ is unable to retrieve a valid watermark from $B_{\mcl U}$ or from the outputs of a surrogate model $H'$ trained on $B_{\mcl U}$.
The objective of this attack is denoted as
\begin{equation}
    \begin{aligned}
        \forall \ b_u \in B_{\mcl U} \cup B'', \mcl R(b_u) \neq w + \epsilon.
    \end{aligned}
\end{equation}
In other words, the goal here is to purge the signal of the watermark inside the container images so that the surrogate model trained on them does not contain the watermark's information.
Thus, the watermarking method is nullified.

Conversely, to counter the overwriting attack, we need a watermarking network that is sufficiently robust so as to be able to retrieve a valid recovered watermark $w'$ under such an attack.
The objective of the defense is denoted as
\begin{equation}
    \begin{aligned}
        \exists \ \mcl R, w' = \mcl R(b_u), \text{ s.t. } w' = w + \epsilon, \forall \ b_u \in B_{\mcl U} \cup B''.
    \end{aligned}
\end{equation}
This objective requires the watermarking method to completely withstand the overwriting attack.

\subsection{The Attack Network}

    
    


\subsubsection{The overwriting attack}
\paragraph{Overview}
As depicted in Figure \ref{fig:usurper}, the overwriting attack aims at the output image set $B'$, which contains the watermark.
A deep steganographic model $\mcl U$ is trained, which consists of an embedding function $\mcl E_{\mcl U}$ and a retrieval function $\mcl R_{\mcl U}$. 
As illustrated in Algorithm \ref{alg:atk_U}, this model is capable of embedding an arbitrary image into another arbitrary image so as to perform an overwriting attack on the given container image set $B''$.
The result is a set of overwritten container images $B_{\mcl U}$, where $w'$ cannot be validly retrieved.

\begin{algorithm}
    \caption{Train the Overwriting Network}\label{alg:atk_U}
    \begin{algorithmic}
        \While{$\mcl L_{\mcl U}$ not converged}
        \State $c_{\mcl U} \gets \mcl E_{\mcl U}(x, c')$ \Comment{Overwrite}
      
        \State $x' \gets \mcl R_{\mcl U}(c_{\mcl U})$ \Comment{Retrieve}
      
        \State $\mcl L_{\mcl U} \gets \mcl L^{\mcl U}_{\mcl E}(c', c_{\mcl U}) + \mcl L^{\mcl U}_{\mcl R}(c_{\mcl U}, x')$ \Comment{Get loss}
      
        \State $\mcl L_{\mcl U}$.back\_propagation() \Comment{Backwards}
    \EndWhile
    \end{algorithmic}
\end{algorithm}

This attack is denoted as
\begin{equation}
    \begin{aligned}
        \mcl E_{\mcl U}(B') = B_{\mcl U}, \ \text{s.t.} \ \mcl R(B_{\mcl U}) \not= w + \epsilon.
    \end{aligned}
\end{equation}
Since the watermark information in $B_{\mcl U}$ is lost, an attacker can train a surrogate model $\mcl H_{\mcl U}$ with $A$ and $B_{\mcl U}$, which is either watermark-free, or it contains a self-made watermark $w_{\mcl U}$.


\paragraph{Loss Functions.} 

The loss function for training $\mcl U$ is defined as
\begin{equation}
    \begin{aligned}
        \mcl L_{\mcl U} = \mcl L_{\mcl E}^{\mcl U} + \mcl L_{\mcl R}^{\mcl U},
    \end{aligned}
    \label{eq:loss_u}
\end{equation}
where $\mcl L_{\mcl E}^{\mcl U}$ and $\mcl L_{\mcl R}^{\mcl U}$ respectively denote the embedding loss and the retrieval loss of $\mcl U$.

$\mcl L_{\mcl E}^{\mcl U}$ is further decomposed into
\begin{equation}
    \begin{aligned}
        \mcl L_{\mcl E}^{\mcl U} = \lambda_{mse} l_{mse} + \lambda_{vgg} l_{vgg} + \lambda_{freq} l_{freq},
    \end{aligned}
    \label{eq:loss_u_e}
\end{equation}
where the $\lambda$s are weight parameters.

$l_{bs}$ is the $L2$ loss between the cover images $C$ and container images $C'$, defined as
\begin{equation}
    \begin{aligned}
        l_{mse} = \sum_{c_i \in C, c_i' \in C'} \frac{1}{N_c} \| c_i - c_i'\|^2,
    \end{aligned}
\end{equation}
where $N_c$ is the total number of pixels.

$l_{vgg}$ denotes the perceptual loss between $C$ and $C'$, defined as
\begin{equation}
    \begin{aligned}
        l_{vgg} = \sum_{c_i \in C, c_i' \in C'} \frac{1}{N_f} \| VGG_k(c_i) - VGG_k(c_i')\|^2,
    \end{aligned}
\end{equation}
where $N_f$ and $VGG_k$ respectively denote the total number of feature neurons and the features extracted at layer $k$.

$l_{freq}$ is the frequency loss \cite{jiang2021focal} between $C$ and $C'$ for controlling consistency in the frequency domain, defined as
\begin{equation}
    \begin{aligned}
        l_{freq} = \sum_{c_i \in C, c_i' \in C'} \frac{1}{N_p} \mcl F(c_i, c_i'),
    \end{aligned}
\end{equation}
where $\mcl F$ and $N_p$ are the focal frequency loss function and the total number of image pairs.

$\mcl L_{\mcl R_{\mcl U}}$ is also further decomposed into
\begin{equation}
    \begin{aligned}
        \mcl L_{\mcl R}^{\mcl U} = \lambda_{mse} l_{mse} + \lambda_{vgg} l_{vgg} + \lambda_{freq} l_{freq},
    \end{aligned}
\end{equation}
where the terms therein are identical to those in $\mcl L_{\mcl E_{\mcl U}}$ but applied to image pairs $(s_i, s_i')$ from the secret images $S$ and the retrieved secret images $S'$.



\subsection{The defense Network}

\begin{figure*}[t!]
    \centering
    \includegraphics[width=.8\textwidth]{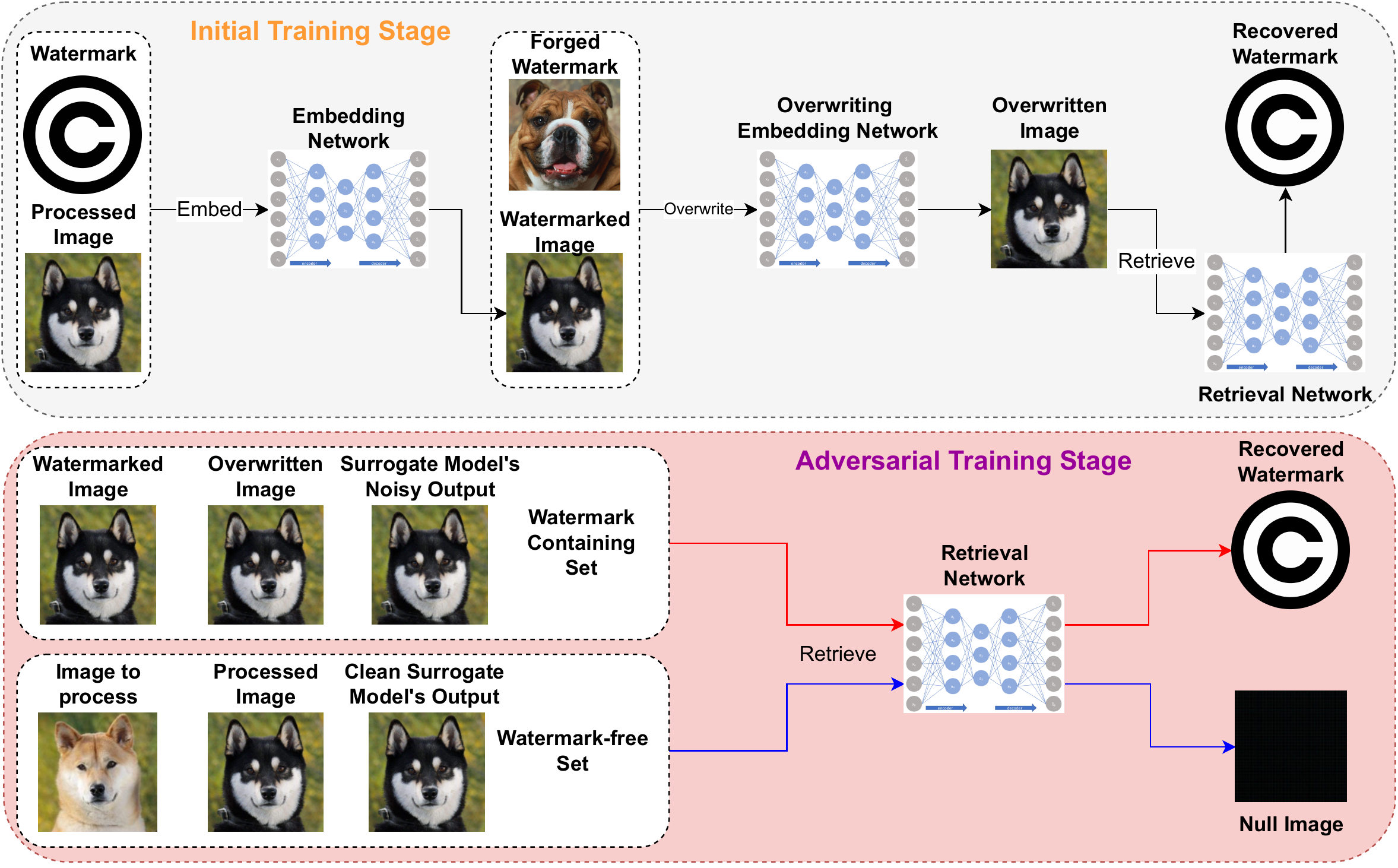}
    \caption{Framework of the Defense Network: 
    The initial training stage is critical for the success of the proposed method, as it involves the concurrent training of both the watermarking and overwriting networks. The retrieval network must be able to extract a valid watermark from the overwritten images generated by the overwriting embedding network. During the adversarial training stage, the retrieval network is further refined through exposure to both watermarked images and watermark-free images. If the retrieval network encounters a watermarked image, it should produce a valid recovered watermark. Conversely, when it encounters a watermark-free image, it should output a null image.
    }
    \label{fig:oathkeeper}
\end{figure*}

\subsubsection{Overview}
To counter the threat of the attack network, we devised an adversarial training framework, i.e., the defense network $\mcl O$, that includes both the watermarking framework $f_{wm}$ and $\mcl U$, and where $f_{wm}$ and $\mcl U$ are each configured as a two-party-mini-max game. 
In short, we set up an adversarial training scheme by training $f_{wm}$ along with $\mcl U$ according to the following settings in Figure \ref{fig:oathkeeper}:

As demonstrated in Algorithm \ref{alg:def_O_init}, the embedding network $\mcl E$ in $f_{wm}$ is initially trained to embed $w$ into $B$ to get $B'$. 
A discriminator network $\mcl D$ then determines whether $B'$ contains a watermark so as to make $\mcl E$ to hide the watermark more covertly.
Meanwhile, the overwriting embedding network $\mcl E_{\mcl U}$ is trained to embed an arbitrary image into another arbitrary image so as to perform an overwriting attack.
$B'$ is then fed to $\mcl E_{\mcl U}$ along with an arbitrary image set $S$ of the same size of $B'$ to yield an overwritten image set $B_{\mcl U}$. 
Lastly, $B'$ and $B_{\mcl U}$ are passed to the retrieval network $\mcl R$ in $f_{wm}$ to retrieve $w'$, and $\mcl R$ is required to produce a null image $w_0$ when it receives watermark-free images from $A$ and $B$.

\begin{algorithm}
    \caption{The Defense Network - Initial Training Stage}
    \label{alg:def_O_init}
    \begin{algorithmic}
        \While{$\mcl L$ not converged}
        \State $B' \gets \mcl E(w, B)$ \Comment{Embed}
      
        \State $B_{\mcl U} \gets \mcl E_{\mcl U}(w_{\mcl U}, B')$ \Comment{Overwrite}
      
        \State $w_0 \gets \mcl R(A; B)$ \Comment{Null retrieval}
      
        \State $w' \gets \mcl R(B'; B_{\mcl U})$ \Comment{Ret. watermark}
      
        \State $\mcl L \gets \mcl L_{\mcl U}(A, B, B', B_{\mcl U}, w, w_{\mcl U}, w_0, w')$ \Comment{Get Loss}
        
        \State $\mcl L \gets \mcl L + \mcl L_{\mcl O}(A, B, B', B_{\mcl U}, w, w_{\mcl U}, w_0, w')$
      
        \State $\mcl L$.back() \Comment{Backwards}
    \EndWhile
    \end{algorithmic}
    
    
\end{algorithm}

At the adversarial training stage, as illustrated in Algorithm \ref{alg:def_O_adv}, only $\mcl R$ is trained for better robustness.
On top of the previous training settings, $\mcl R$ is further forced to retrieve a watermark from the noisy output $B''$ generated by the surrogate model $H'$.
Meanwhile, a clean surrogate model $H_0$ is trained to produce clean output $B_0$, which boosts the specificity of $\mcl R$, where $\mcl R$.
Further, $\mcl R$ must also retrieve a null image when it receives $B_0$.
This solves an intractable problem that we encountered in the experiments, which is further discussed in Section \ref{sect:problem}.

\begin{algorithm}
    \caption{The defense Network - Adversarial Training Stage}
    \label{alg:def_O_adv}
    \begin{algorithmic}
        \While{$\mcl L_{\mcl O}$ not converged}
            \State $B_0 \gets \mcl H_0(A)$
              
            \State $B'' \gets \mcl H'(A)$
              
            \State $w_0 \gets \mcl R(A; B; B_0)$\Comment{Null Retrieval}
              
            \State $w' \gets \mcl R(B'; B''; B_{\mcl U})$\Comment{Watermark Retrieval}
              
            \State $\mcl L_{\mcl O} = \mcl L_{\mcl O}(A, B, B', B'', w, w_0, w')$\Comment{Get Loss}
              
            \State $\mcl L_{\mcl O}$.back()\Comment{Backwards}
        \EndWhile
    \end{algorithmic}
    
    
      
\end{algorithm}

The two-party-mini-max game is defined as
\begin{equation}
    \begin{aligned}
        & \qquad \qquad \qquad \underset{\mcl E, \mcl R}{\min} \ \underset{\mcl E_{\mcl U}}{\max} \ \mcl L(\mcl E, \mcl R, \mcl E_{\mcl U}) =\\
        & \bigg( 
            \mbb E \big[ \sum_{b_i \in B, s_i \in S} \frac{1}{N_c} \big\| 
                \mcl R \big( 
                    \mcl E_{\mcl U}(
                        \mcl E(b_i, w), 
                        s_i) 
                    \big) 
                - w \big\|^2 \big]
            \bigg),
    \end{aligned}
\end{equation}
where $\mcl E_{\mcl U}$ mostly benefits when $\mcl R$ cannot retrieve a valid $w'$. 
Additionally, $\mcl E$ and $\mcl R$ get the most bonuses when a $w'$ that is infinitely close to $w$ is retrieved.

\subsubsection{Loss Functions} 

The loss function for training the defense network is defined as 
\begin{equation}
    \begin{aligned}
        \mcl L = \mcl L_{\mcl U} + \mcl L_{\mcl O}
    \end{aligned}
\end{equation}
where $\mcl L_{\mcl U}$ and $\mcl L_{\mcl O}$ respectively denote the loss of training the overwriting network and the watermarking part of the defense network.
Similar to Equation \ref{eq:loss_u}, $\mcl L_{\mcl U}$ here is defined as
\begin{equation}
    \begin{aligned}
        \mcl L_{\mcl U} = \mcl L_{\mcl E}^{\mcl U} + \mcl L_{\mcl R}^{\mcl U} + l_{\mcl U}.
    \end{aligned}
\end{equation}
The extra term $l_{\mcl U}$ denotes the adversarial overwriting loss that attempts to make $\mcl R$ retrieve a blank image $w_0$ from $B_{\mcl U}$.
This is defined as
\begin{equation}
    \begin{aligned}
        l_{\mcl U} = \sum_{b_i \in B_{\mcl U}} \frac{1}{N_c} \| \mcl R(b_i) - w_0 \|^2.
    \end{aligned}
\end{equation}

$\mcl L_{\mcl O}$ is then further decomposed into
\begin{equation}
    \begin{aligned}
        \mcl L_{\mcl O} = \mcl L^{\mcl O}_{\mcl E} + \mcl L^{\mcl O}_{\mcl R} + \mcl L^{\mcl O}_{\mcl D} + l_{\mcl O}
    \end{aligned}
\end{equation}
where the terms represent the loss of training the embedding network, the retrieval network, the discriminator, and the defense adversarial loss.
Further, $\mcl L_{\mcl E}^{\mcl O}$ comprises the following terms:
\begin{equation}
    \begin{aligned}
        \mcl L_{\mcl E}^{\mcl O} = \lambda_{mse} l_{mse} + \lambda_{freq} l_{freq} + \lambda_{vgg} l_{vgg} + \lambda_{adv} l_{adv},
    \end{aligned}
\end{equation}
where the former three losses are identical to those appearing in Equation \ref{eq:loss_u_e}.
The last term $l_{adv}$ represents the adversarial loss against the discriminator network, defined as
\begin{equation}
    \begin{aligned}
        l_{adv} = \mbb E_{b_i' \in B'} \big[ \log(\mcl D(b_i')) \big].
    \end{aligned}
\end{equation}
The goal is to make the embedding network produce container images that cannot be detected by the discriminator network.

$\mcl L_{\mcl R}$ is decomposed into
\begin{equation}
    \begin{aligned}
        \mcl L_{\mcl R} = \lambda_{wm} l_{wm} + \lambda_{clean} l_{clean} + \lambda_{cst} l_{cst},
    \end{aligned}
\end{equation}
where the $\lambda$s are weight parameters.
$\mcl L_{wm}$ denotes watermark retrieval loss
\begin{equation}
    \begin{aligned}
        l_{wm} = \sum_{b_i' \in B'} \frac{1}{N_c} \| \mcl R(b_i') - w\|^2 + \sum_{b_i'' \in B''} \frac{1}{N_c} \| \mcl R(b_i'') - w\|^2.
    \end{aligned}
\end{equation}
$\mcl L_{cln}$ represents the blank extraction loss for guiding $\mcl E$ to extract only blank images from images not possessing watermark information, denoted as
\begin{equation}
    \begin{aligned}
        l_{clean} = \sum_{a_i \in A} \frac{1}{N_c} \| \mcl R(a_i) - w_0\| + \sum_{b_i \in B} \frac{1}{N_c} \| \mcl R(b_i) - w_0\|,
    \end{aligned}
\end{equation}
where $w_0$ is a blank image.
Lastly, $\mcl L_{cst}$ is the consistency loss for ensuring that the watermarks extracted from different images are consistent, denoted as
\begin{equation}
    \begin{aligned}
        l_{cst} = \sum_{x, y \in B' \bigcup B''} \| \mcl R(x) - \mcl R(y) \|^2
    \end{aligned}
\end{equation}


$l_{\mcl O}$ stands for the defense adversarial loss that guarantees that $\mcl R$ can retrieve $w' = w + \epsilon$ from the overwritten images $B_{\mcl U}$, defined as
\begin{equation}
    \begin{aligned}
        l_{\mcl O} = \sum_{b_i \in B_{\mcl U}} \frac{1}{N_c} \| \mcl R(b_i) - w \|^2.
    \end{aligned}
\end{equation}




\subsection{Discussion}

In our defense framework, the overwriting network is trained in tandem with the watermarking network to form the defense network. The purpose of the overwriting network is to overwrite the original watermark with a forged watermark, creating an adversarial relationship between the two. The retrieval network of the watermarking network must then be able to retrieve the original watermark from the overwritten images, as demonstrated in previous work \cite{baluja2019hiding}.

As the two embedding networks embed the original and forged watermarks into the same container image in parallel, both secret images are preserved within the container image. This is because, as shown in \cite{baluja2019hiding}, it is possible to embed multiple secret images into one cover image, albeit with a higher perceptual quality loss.
Our experiments show that without proper adversarial training, the watermarking network is unable to retrieve a valid watermark. Thus, our adversarial training scheme is a crucial component of the defense framework.



\section{Experiment}
\label{sect:exp}

\subsection{Experimental Setup}



\subsubsection{Dataset}
Two datasets were used to train the image-processing surrogate model: the de-raining dataset from \cite{zhang2018density} and an 8-bit image dataset generated via the algorithm in \cite{zou2021stylized}.
The de-raining dataset is public available, while the 8-bit image dataset is generated using the images in ImageNet datset.
The goal of the first task was to remove rain drops from the images. 
The second task was to transform an input image into an 8-bit style artwork.
With each task, we split the dataset into two subsets: a training set of $4000$ images and a test set of 1000 images. 
All the images were resized to $256 \times 256$ for training.

We also took samples from the ImageNet dataset to train $\mcl U$. 
Here, there were $40,000$ images in the training set and $10,000$ images in the test set. 
Each image was greater than $256 \times 256$ in size so we randomly cropped the images down to $256 \times 256$ so as to enhance $\mcl U$'s robustness.

\subsubsection{Implementation details}
$\mcl E$'s network structure follows UNet \cite{ronneberger2015u}. 
UNet has considerably high performance with reference to translation based and semantic segmentation tasks, so the network produces good results when there are close connections between the inputs and the outputs. 
CEILNet \cite{fan2017generic} was used as the model for $\mcl R$, which is believed to work well when the inputs somewhat differ from the outputs.
Patch-GAN \cite{isola2017image} was used for $\mcl D$.

In terms of the overwriting network, UNet \cite{ronneberger2015u} was once more used as the network structure for $\mcl E_{\mcl U}$.
For $\mcl R_{\mcl U}$, we simply used stacks of convolutional layers, as the critical point lies in the overwriting procedure, and, here, the embedding network plays the more crucial role.
Moreover, there is no discriminator in this training process. Lastly, the defense network comprised the watermarking network and $\mcl U$. 
Together, they form $\mcl O$.

\subsubsection{Evaluation metrics}
We chose PSNR and SSIM \cite{hore2010image} to evaluate the visual quality of the container image in comparison to its corresponding cover image. 
Additionally, we used normalized cross correlation (NCC) to measure whether a retrieved watermark was valid. 
If the NCC between a retrieved watermark and the original watermark was greater than $0.95$, the retrieved watermark was considered to be legitimate. 
NCC is defined as
\begin{equation}
    \begin{aligned}
        NCC = \frac{\langle \mcl R (b_i'), w \rangle}{\| \mcl R (b_i') \| \cdot \| w\|}
    \end{aligned}
\end{equation}
where $\langle \cdot , \cdot \rangle$ denotes the inner product, and $\| \cdot \|$ denotes the L2 norm.
The success rate $SR$ is defined as the ratio of successfully retrieved watermarks from a certain amount of container images.
\subsection{Baseline Reproduction}

\subsubsection{Training the watermarking network}
First, we reproduced Zhang's method \cite{zhang2020model} as our experimental baseline.
When training the watermarking networks $\mcl E$, $\mcl R$, and $\mcl D$, we set the initial learning rate of the Adam optimizer to 0.001.
Here, the goal was to equip the watermarking network with the ability to embed a fixed watermark into an arbitrary image, and to retrieve the watermark from the container image.
Therefore, we trained the network on the ``ImageNet" training dataset, where 1 epoch contains 40,000 images.
The images were randomly cropped down to $256 \times 256$ so as to increase the randomness of the input data.
We set the batch size to 10, which means that the model ran through 10 images in one iteration. 
If there was no loss descent within 4000 iterations, we decreased the learning rate by 0.2.
All $\lambda$s were set to 1 except for $\lambda_{3} = 0.01$, which is the weight parameter of the adversarial loss.

\begin{figure}[t!]
    \centering
    
    \begin{subfigure}[b]{.24\textwidth}
        \centering
        \includegraphics[width=\textwidth]{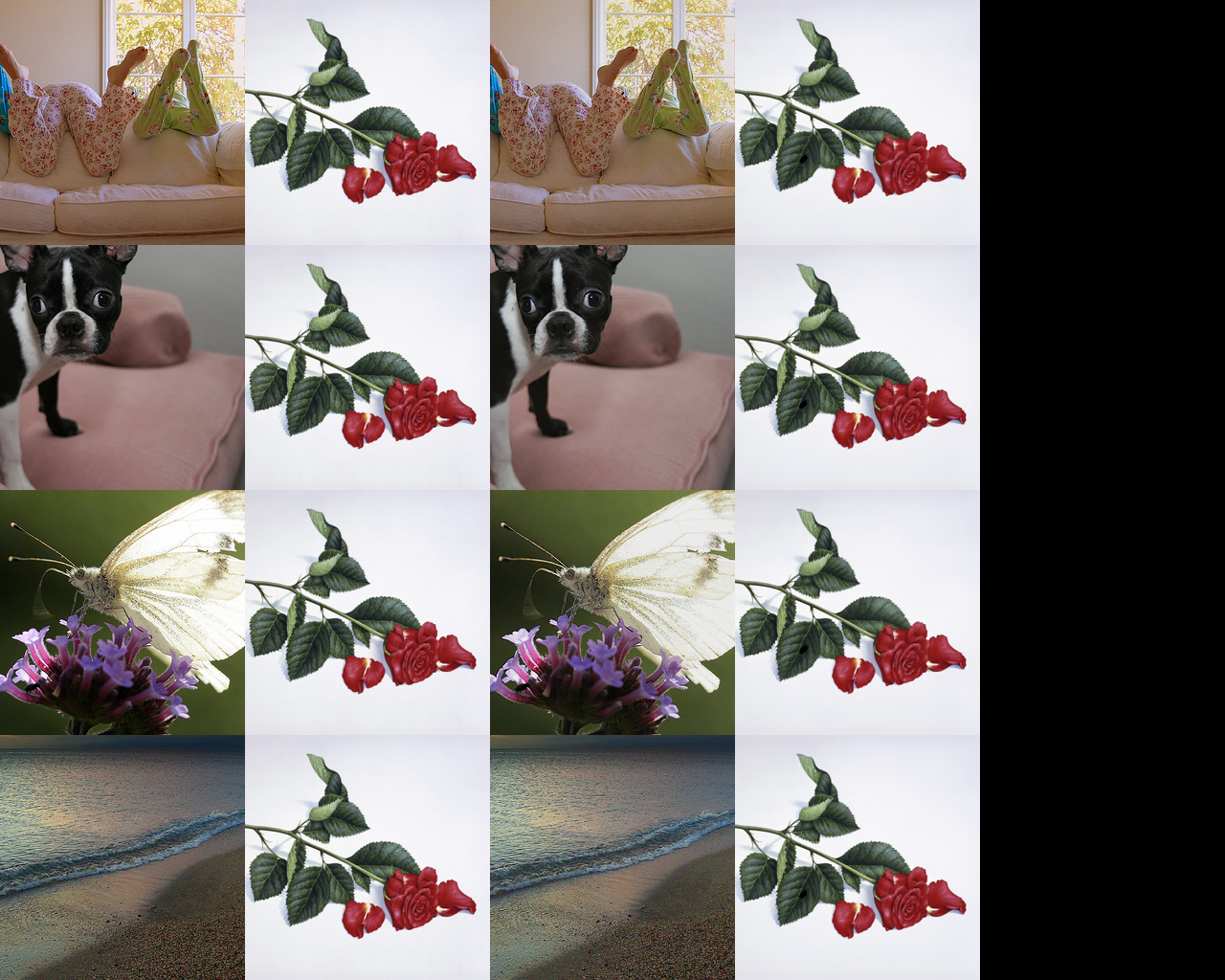}
        \caption{Flower}
    \end{subfigure}
    \hfill
    \begin{subfigure}[b]{.24\textwidth}
        \centering
        \includegraphics[width=\textwidth]{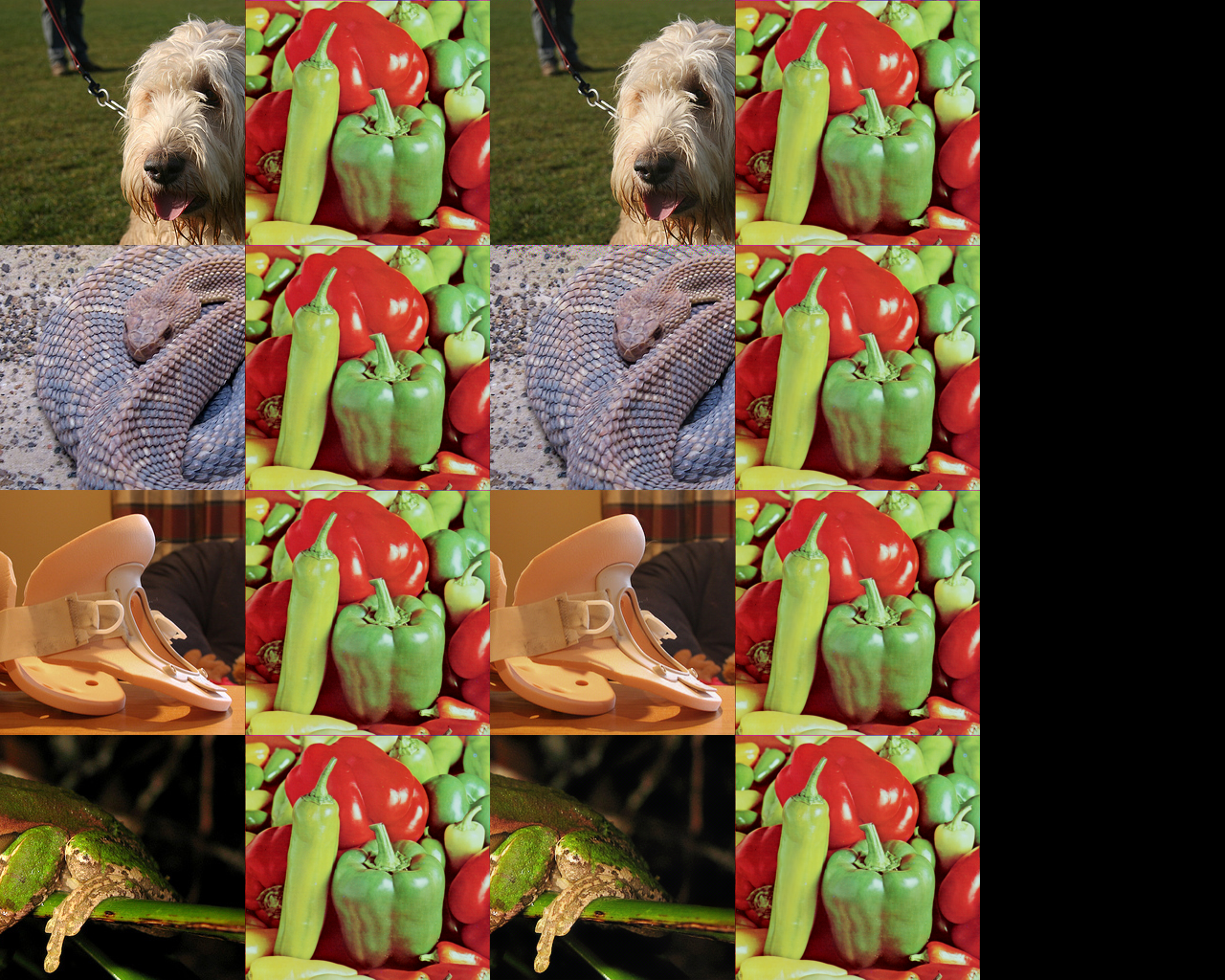}
        \caption{Pepper}
    \end{subfigure}
    
    \caption{Test Results of the Watermarking Network}
    \label{fig:test_wmer_res}
\end{figure}

Figure \ref{fig:test_wmer_res} depicts the test results of the two trained models.
Each row of the two images is an instance.
From left to right, each column represents the cover images $c$, the secret images $s$, the container images $c'$, the retrieved secret images $s'$, and the null images retrieved from watermark-free images.
From the results, it is clear that the watermarking network was able to complete both the embedding and the retrieval tasks quite well. 
Further, a pure black image was guaranteed when the input contained no hidden content.
Here, our settings slightly differ with Zhang's method. 
Zhang set the null image to pure white, but for ease of reading, we set the null image to black.

\subsubsection{The adversarial stage}
This stage was designed to enhance the robustness of the retrieval network by training it to retrieve the watermark from the surrogate model's processed images.
The retrieval will fail if this step is not conducted, because the retrieval network does not meet any noisy samples from the surrogate model.
However, because of a problem we discovered (discussed in Section \ref{subsect:atk}), we made a change to this stage of the process where we involved the outputs from a watermark-free surrogate model in the fine-tuning process.

To train the surrogate models, we used the de-raining and the 8-bit datasets in two parallel experiments.
The paired processed images were watermarked by the watermarking network.
By training the surrogate models this way, we forced the surrogate models to overfit the watermark signals hidden in the processed images, such that every output from the surrogate models carried the watermark signal.
Here, the batch size was set to $20$ and the number of epochs was set to $50$ based on some previous experiments. 
The initial learning rate was set to $0.001$, and then decayed by $0.2$ if the loss remained unchanged for 5 epochs.
Additionally, we use the same settings to train a watermark-free surrogate model with the watermark-free datasets.

After training the surrogate models, we used them to produce noisy watermarked images and the watermark-free images, which were then fed into the retrieval network $\mcl R$. 
In the adversarial training state, the hyperparameters remained the same as those in training the watermarking network for updating $\mcl R$.
However, we reset the learning rate back to $0.001$ so as to let the trained network escape reach of the local minimum.
The fine-tuning process lasted for $50$ epochs.

\begin{figure}[t!]
    \centering
    
    \begin{subfigure}[b]{.24\textwidth}
        \centering
        \includegraphics[width=\textwidth]{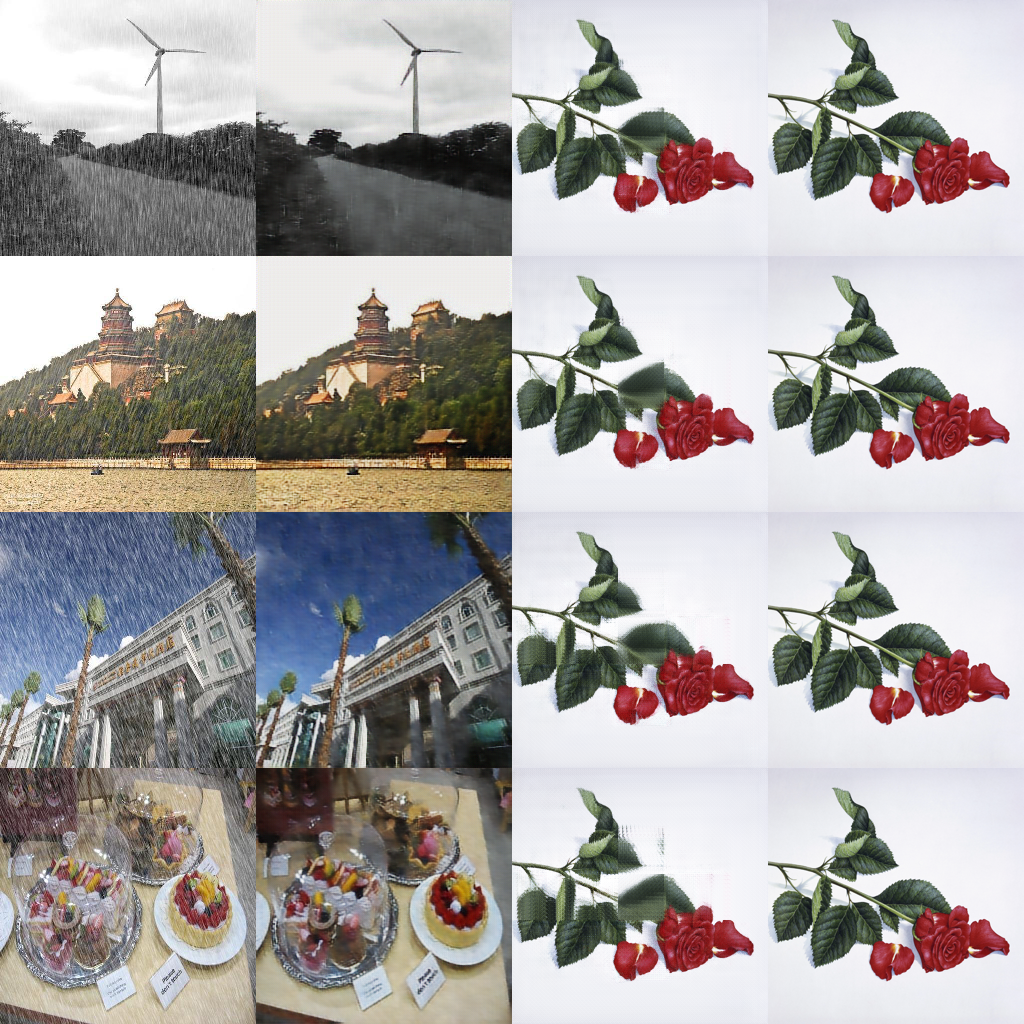}
        \caption{De-raining + Flower}
    \end{subfigure}
    \hfill
    \begin{subfigure}[b]{.23\textwidth}
        \centering
        \includegraphics[width=\textwidth]{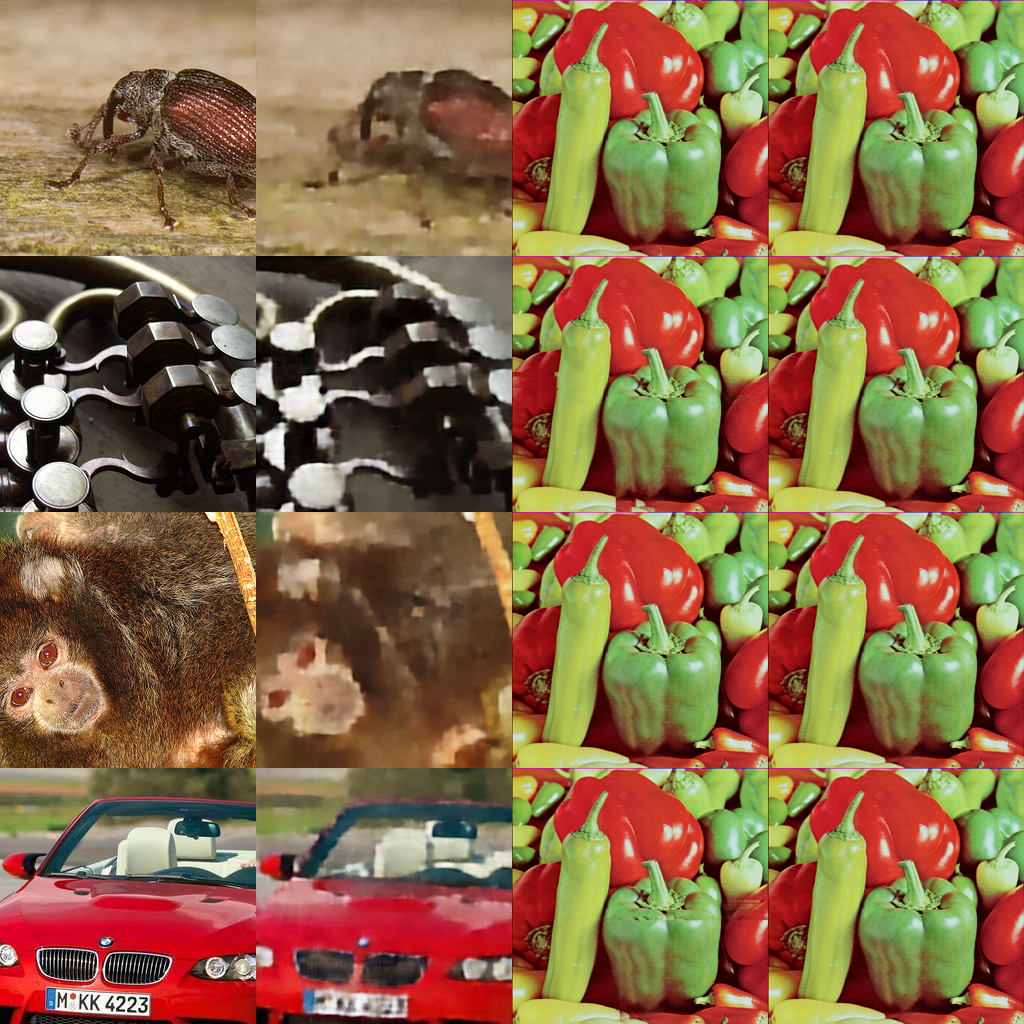}
        \caption{8bit + Pepper}
    \end{subfigure}
    
    \caption{Watermarks Retrieved from the Surrogate Models' Output}
    \label{fig:org_meth_res}
\end{figure}

As a result, the fine-tuned retrieval network was able to retrieve the watermark from both the watermarked image $B'$ and the surrogate model's outputs $B''$. 
The results are visually presented in Figure \ref{fig:org_meth_res}.
Details of the results are listed in Table \ref{tab:res_org}. 

\begin{table}[t!]
  \centering
  \scriptsize
  \caption{Original Method Results}
  \renewcommand\arraystretch{1.5}
  \setlength{\tabcolsep}{2mm}{
  
    \begin{tabular}{l|ccccccc}
    \toprule
     Condition/Metric & \textbf{PSNR} & \textbf{SSIM} & \textbf{NCC} & \textbf{SR}(\%) \\
    \midrule
    \textbf{De-raining $\times$ $\mcl W$} & 30.49 & 0.8688 & 0.9992  & 100 \\
    \textbf{De-raining $\times$ $\mcl W$ $\times$ UNet} & /  & /  & 0.9974  & 100 \\
    \textbf{De-raining $\times$ $\mcl W$ $\times$ Res16} & /  & /  & 0.9877  & 100 \\
    \textbf{8-bit $\times$ $\mcl W$} & 32.89  & 0.8739  & 0.9999  & 100 \\
    \textbf{8-Bit $\times$ $\mcl W$ $\times$ UNet} & /  & /  & 0.9985  & 100 \\
    \textbf{8-Bit $\times$ $\mcl W$ $\times$ Res16} & /  & /  & 0.9910  & 100 \\
    \bottomrule
    \end{tabular}}%
  \label{tab:res_org}%
\end{table}%

With each dataset, we conducted three parallel experiments: one using UNet, one using a residue network with $16$ blocks (Res16), and one performed directly on the watermarked image $B'$.
PSNR and SSIM were used to measure the quality of the container image $c'$ compared to its corresponding cover image $c$.
NCC and SR were only used to validate the watermark retrieval.
Remarkably, the success rate of the watermark retrieval reached $100\%$ in each experiment, which firmly verifies the efficacy of Zhang's method.

\subsection{Attacks}
\label{subsect:atk}

\begin{figure}[t!]
    \centering
    \includegraphics[width=.3\textwidth]{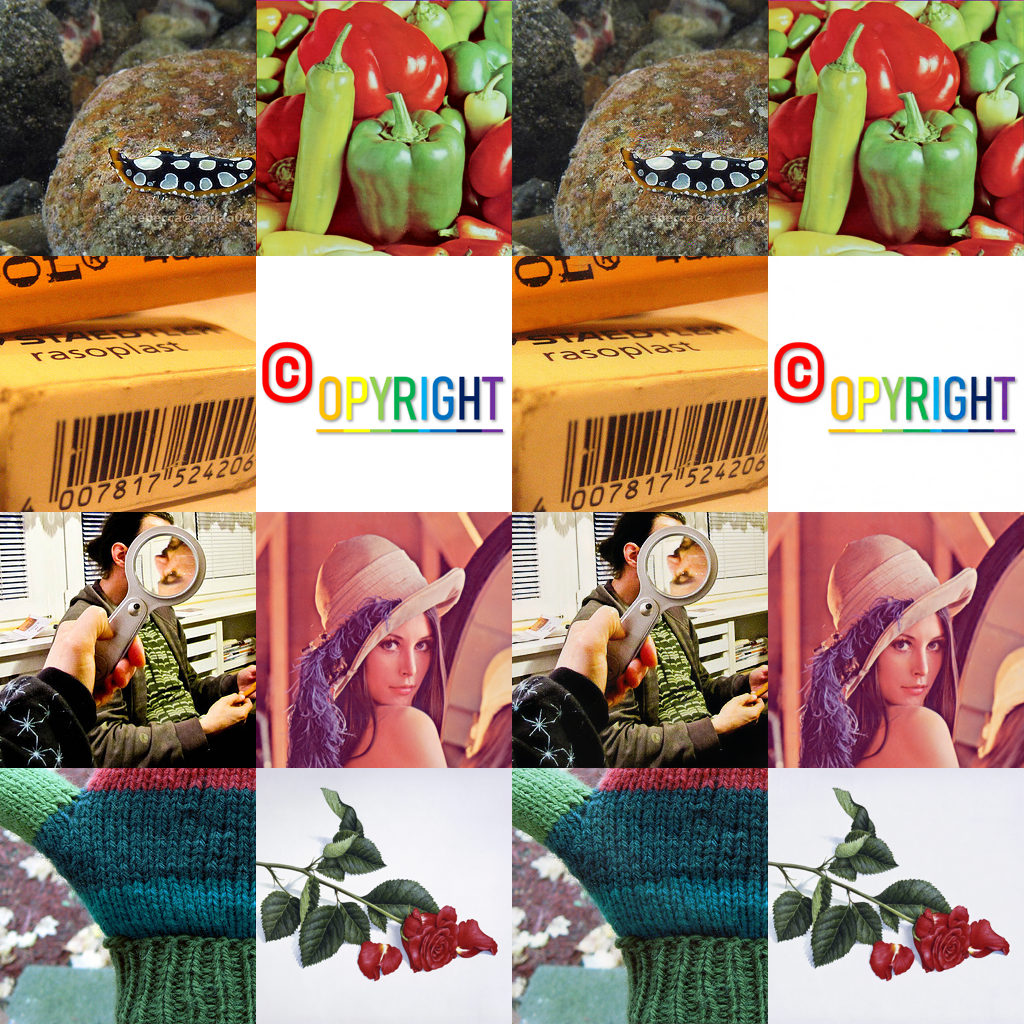}
    \caption{Test Results of the Overwriting Network}
    \label{fig:test_u}
\end{figure}

We trained our overwriting network with the Adam optimizer on the ImageNet training set. 
The learning rate and batch size were set to $0.001$ and $20$.
We decreased the learning rate by $0.2$ if there was no loss decrease within $2,000$ iterations.
After $20$ epochs, there was no significant descent loss, so we set the number of epochs to $30$.
The $\lambda$s were all set to $1$.
The cover images were randomly cropped into size of $256 \times 256$ so as to increase randomness of the input data and, in turn, enhance the robustness of the overwriting network.
Further, the overwriting network was trained to embed one of four selected watermarks: ``flower", ``copyright", ``lena", or ``pepper", into an arbitrary image, and then retrieve the watermark.
The effect is depicted in Figure \ref{fig:test_u}, where each row is an instance. 
From left to right, each column respectively represents the cover images $c$, the secret images $s$, the container images $c'$, and the recovered secret images $s'$.


\begin{figure}[t!]
    \centering
    
    \begin{subfigure}[b]{0.24\textwidth}
        \centering
        \includegraphics[width=\textwidth]{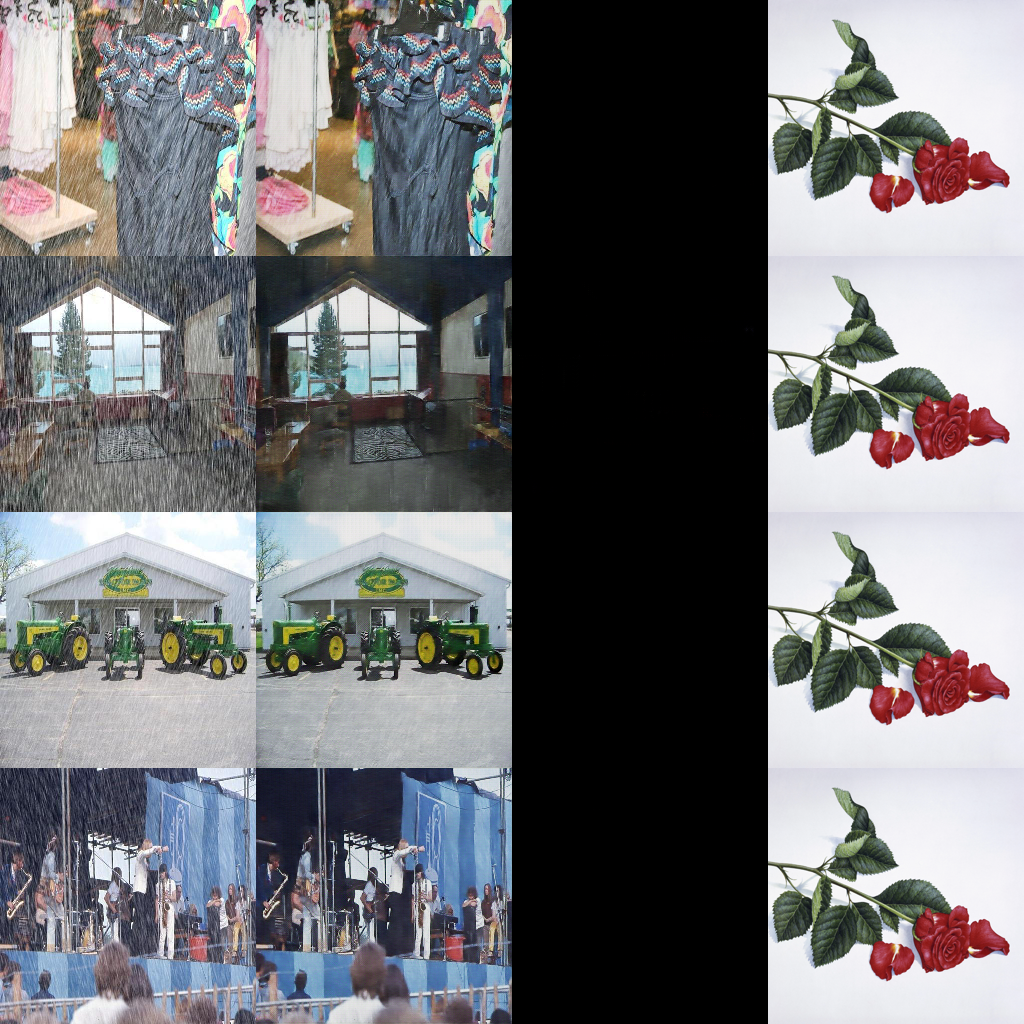}
        \caption{De-raining + Flower}
    \end{subfigure}
    \hfill
    \begin{subfigure}[b]{0.24\textwidth}
        \centering
        \includegraphics[width=\textwidth]{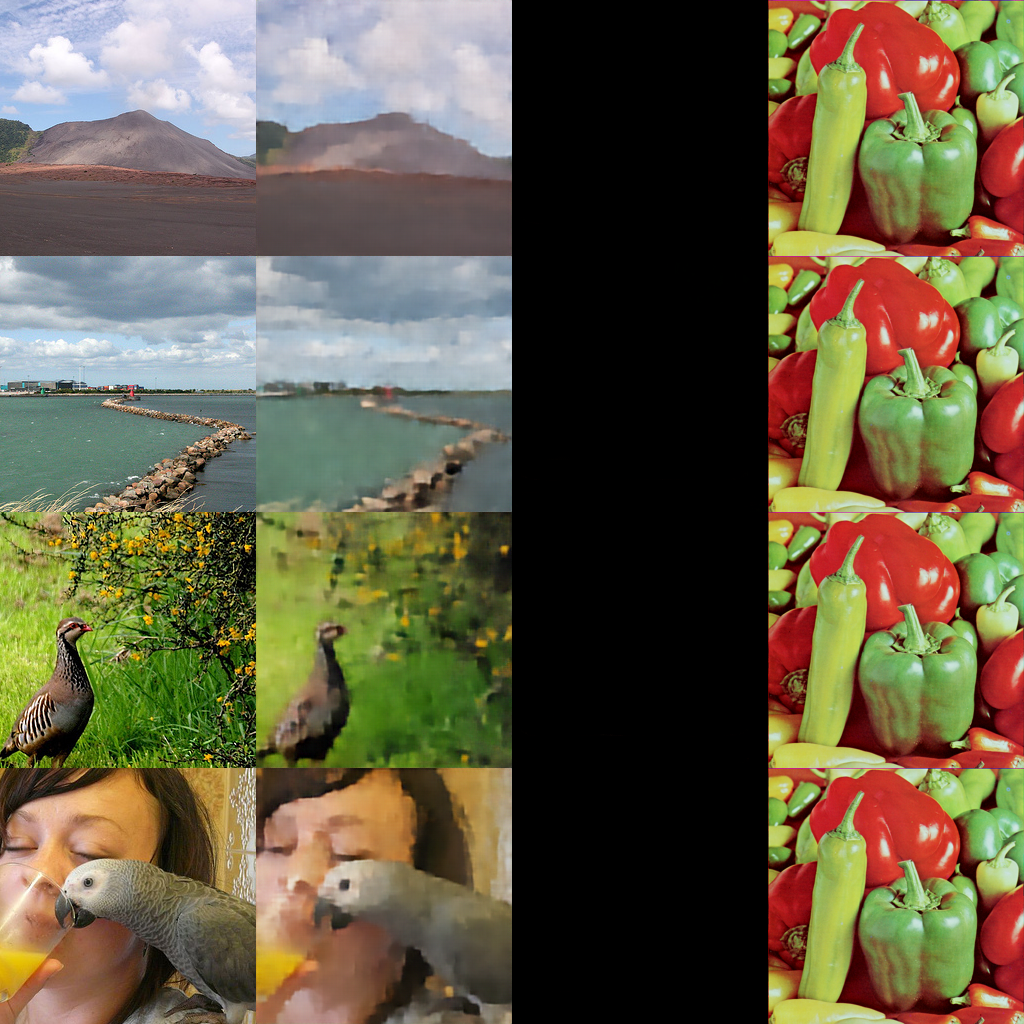}
        \caption{8bit + Pepper}
    \end{subfigure}

    \caption{Results of the Overwriting Attack}
    \label{fig:usurper_atk_res}
\end{figure}

After having the trained overwriting network, we launched an attack on the watermarked images.
Generally, the watermarked images $B'$ were overwritten with another watermark so as to prevent the watermark from being retrieved.
The direct effect of the attack is depicted in Figure \ref{fig:usurper_atk_res}, where each row is an instance. 
From left to right, each column respectively represents the cover images $c$, the secret images $s$, the container images $c'$, and the retrieved secret images $s'$.
Table \ref{tab:res_atk} lists the results of the visual quality test for the container image and watermark retrieval under various conditions, namely, different combinations of surrogate model types and datasets.
Each value is the average number of the item derived from 100 randomly selected images.
We performed three experiments with each of the two tasks, i.e., a direct attack on $B'$, an attack on the UNet surrogate model, and another on the Res16 surrogate model.

Compared to the watermarked images $B'$, the quality of the attacked image $B_{\mcl U}$ decreased slightly.
However, the quality loss was still negligible to human eye.
Success rates across all the experiments were no greater than 10\%, and half of them reached 0\%, which proves the efficacy of our attack.
Notably, the success rate of the watermark retrieval with the Res16 surrogate model on both tasks was higher than the others, which is an interesting phenomenon.

\begin{table}[t!]
  \centering
  \scriptsize
  \caption{Attack Results}
  \renewcommand\arraystretch{1.5}
  \setlength{\tabcolsep}{2mm}{
  
    \begin{tabular}{l|ccccccc}
    \toprule
     Condition/Metric & \textbf{PSNR} & \textbf{SSIM} & \textbf{NCC} & \textbf{SR}(\%) \\
    \midrule
    \textbf{De-raining $\times$ $\mcl W$ $\times$ $\mcl U$} & 27.24 & 0.8031 & 0.0565  & 4 \\
    \textbf{De-raining $\times$ $\mcl W$ $\times$ $\mcl U$ $\times$ UNet} & / & / & 0.0109 & 0 \\
    \textbf{De-raining $\times$ $\mcl W$ $\times$ $\mcl U$ $\times$ Res16} & / & / & 0.1527  & 10 \\
    \textbf{8-bit $\times$ $\mcl W$ $\times$ $\mcl U$} & 31.91 & 0.6061 & 0.2968 & 0 \\
    \textbf{8-Bit $\times$ $\mcl W$ $\times$ $\mcl U$ $\times$ Unet} & / & / & 0.0678 & 0 \\
    \textbf{8-Bit $\times$ $\mcl W$ $\times$ $\mcl U$ $\times$ Res16} & / & / & 0.2248 & 5 \\
    \bottomrule
    \end{tabular}}%
  \label{tab:res_atk}%
\end{table}%

\subsubsection{The overfitting problem in the retrieval network.}
\label{sect:problem}

\begin{figure*}[t!]
    \centering
    \includegraphics[width=.8\textwidth]{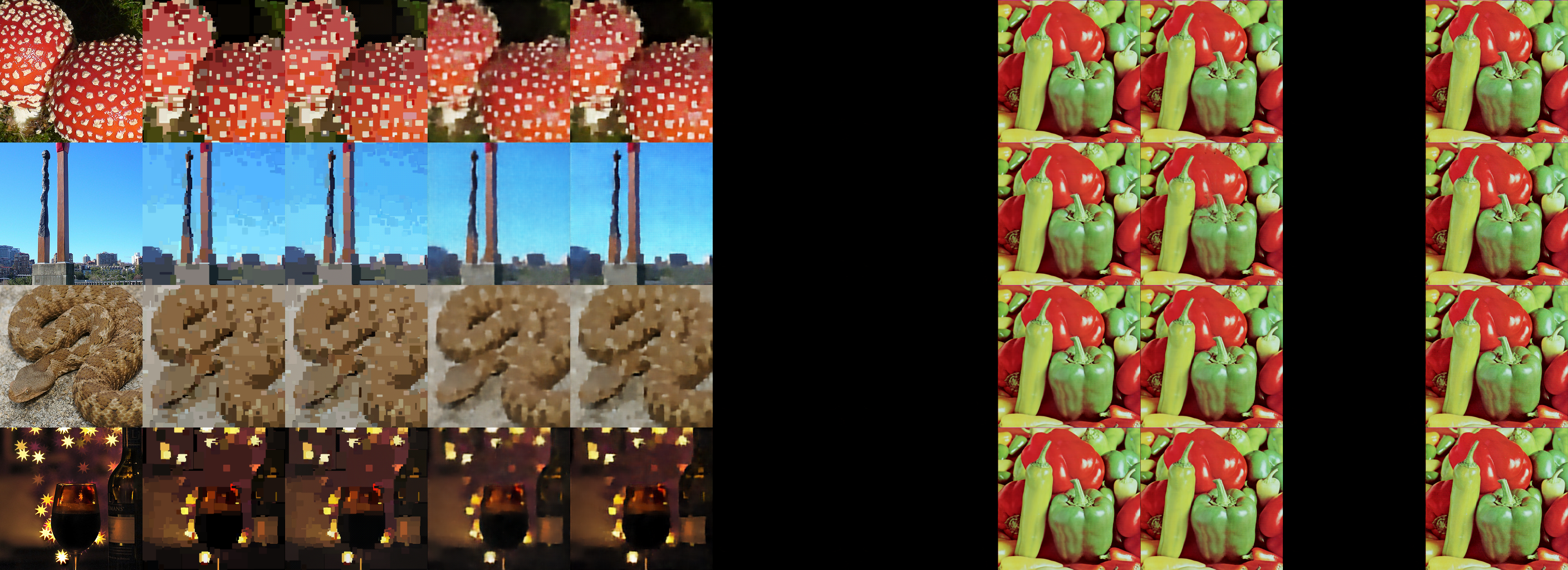}
    \caption{Training Process of the defense Network:
    From left to right: 
    the images to process $A$, 
    the processed images $B$, 
    the watermarked processed images $B'$, 
    the overwritten watermarked processed images $B_{\mcl U}$, 
    the overwritten watermark-free processed images $B_{\mcl U}'$,
    the null images retrieved from $A$ and $B$, 
    the watermarks retrieved from $B'$ and $B_{\mcl U}$,
    the null images retrieved from $B_{\mcl U}'$,
    and the watermark image $w$.
    }
    \label{fig:train_o}
\end{figure*}

\begin{figure}[t!]
    \centering
    \includegraphics[width=.3\textwidth]{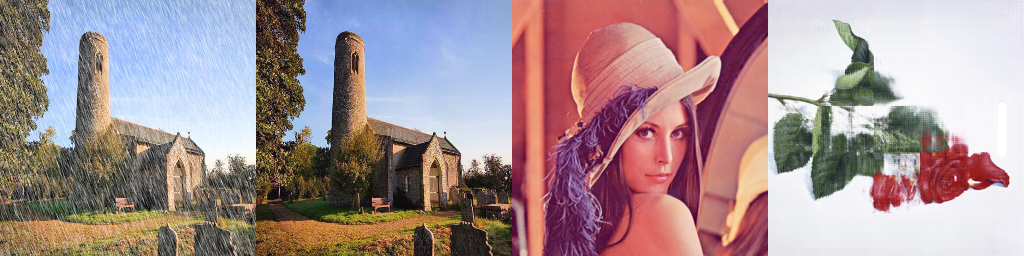}
    \caption{Overfitting Phenomenom:
    From left to right, the images depict the rainy image to process, the watermarked image from the overwriting network, the overwriting watermark, and the retrieved watermark from the second image by the retrieval network.
    The watermark can be retrieved from any container image that has some steganographic content.}
    \label{fig:overfitting}
\end{figure}

In the attack, we also tried to use the fine-tuned retrieval network to extract watermarks from the images that were only processed by the overwriting network.
In other words, we tried to extract watermarks from images that did not contain a watermark signal embedded by the watermarking network.
Under these circumstances, the retrieval network was still able to retrieve the watermark with decreased quality as demonstrated in Figure \ref{fig:overfitting}. 
This indicated that, during the fine-tuning, the retrieval network was tuned to output a pre-defined watermark if there was any secret image signal in the container images, regardless of what exactly the signal represented.

Though this method can withstand an overwriting attack by this overfitting phenomenon, the phenomenon is harmful to this method. 
This is because the watermarking scheme is nullified if a valid watermark can be retrieved from any container image that does not contain the corresponding watermark information.

We managed to overcome this problem with a fairly simple manoeuvre.
We trained a watermark-free surrogate model, and then, we added its output images into the adversarial stage of fine-tuning the retrieval network.
The retrieval network was therefore made to differentiate the outputs of the watermark-free surrogate model from those of the watermarked surrogate model, and output the null images correspondingly.
This extra step successfully mitigates this problem.

\subsection{defenses}

Lastly, we trained the defense network with the same hyperparameters as above.
The main idea was to concurrently train a watermarking network and an overwriting network, and to make the retrieval network retrieve the watermark from the overwritten container image.
Meanwhile, as the adversary, the overwriting network attempts to overwrite the watermark within the container image so that the retrieval network will only yield null images.
Figure \ref{fig:train_o} directly illustrates the training process, where the defense network is trained to embed the watermark into an arbitrary image, and retrieve the watermark from the container and overwritten images.
Further, the retrieval network must generate a null image if there is not the embedded watermark signal in the input.

The settings in the fine-tuning stage were almost the same as for the watermarking network's adversarial stage.
Additionally, the overwriting network also participated in this stage so as to force the retrieval network to produce either the watermark or a null image when it encounters the overwritten container image or the container image generated only by the overwriting network.

Finally, we tested the defense network on two datasets with different settings.
Table \ref{tab:res_def} shows the test results. 
As shown, the container images generated by the defense network have a better visual quality than those generated by the watermarking network.
Among the watermarking retrieval tests, all success rates reached 100\% except for the direct overwriting attack on the 8-bit datasets, which verifies the efficacy of our defense method.

\begin{table}[t!]
  \centering
  \scriptsize
  \caption{Watermark Retrieval Results Comparison among Defenses}
  \renewcommand\arraystretch{1.5}
  \setlength{\tabcolsep}{2mm}{
  
    \begin{tabular}{l|ccccccc}
    \toprule
     Condition/Metric & \textbf{PSNR} & \textbf{SSIM} & \textbf{NCC} & \textbf{SR}(\%) \\
    \midrule
    \textbf{De-raining $\times$ $\mcl O$} & 34.07 & 0.9022 & 0.9997  & 100 \\
    \textbf{De-raining $\times$ $\mcl O$ $\times$ $\mcl U$} & / & / & 0.9924 & 100 \\
    \textbf{De-raining $\times$ $\mcl O$ $\times$ $\mcl U$ $\times$ UNet} & / & / & 0.9915 & 100 \\
    \textbf{De-raining $\times$ $\mcl O$ $\times$ $\mcl U$ $\times$ Res16} & / & / & 0.9914 & 100 \\
    \textbf{8-bit $\times$ $\mcl O$} & 34.54 & 0.8796 & 0.9998 & 100 \\
    \textbf{8-bit $\times$ $\mcl O$ $\times$ $\mcl U$} & / & / & 0.9040 & 0.81 \\
    \textbf{8-bit $\times$ $\mcl O$ $\times$ $\mcl U$ $\times$ UNet} & / & / & 0.9991 & 100 \\
    \textbf{8-bit $\times$ $\mcl O$ $\times$ $\mcl U$ $\times$ Res16} & / & / & 0.9982 & 100 \\
    \bottomrule
    \end{tabular}}%
  \label{tab:res_def}%
\end{table}%



\section{Discussion}
\label{sect:dis}



\subsection{Analysis of the Overwriting Attack}

\begin{figure*}[t!]
    \centering
    
    \begin{subfigure}[b]{.45\textwidth}
        \includegraphics[width=\textwidth]{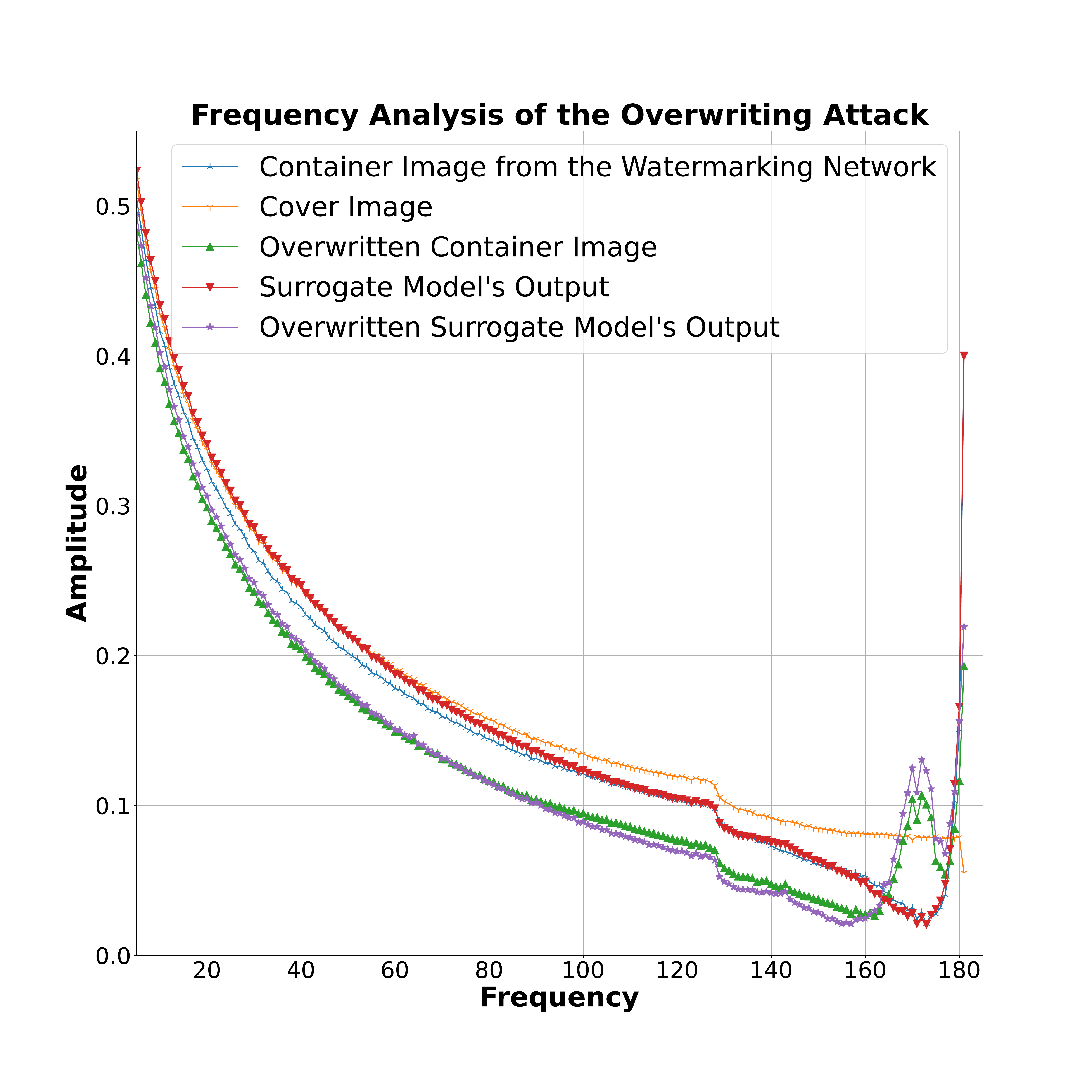}
        \caption{The Overwriting Attack}
        \label{fig:freq_als_atk}
    \end{subfigure}
    \hfill
    \begin{subfigure}[b]{.45\textwidth}
        \centering
        \includegraphics[width=\textwidth]{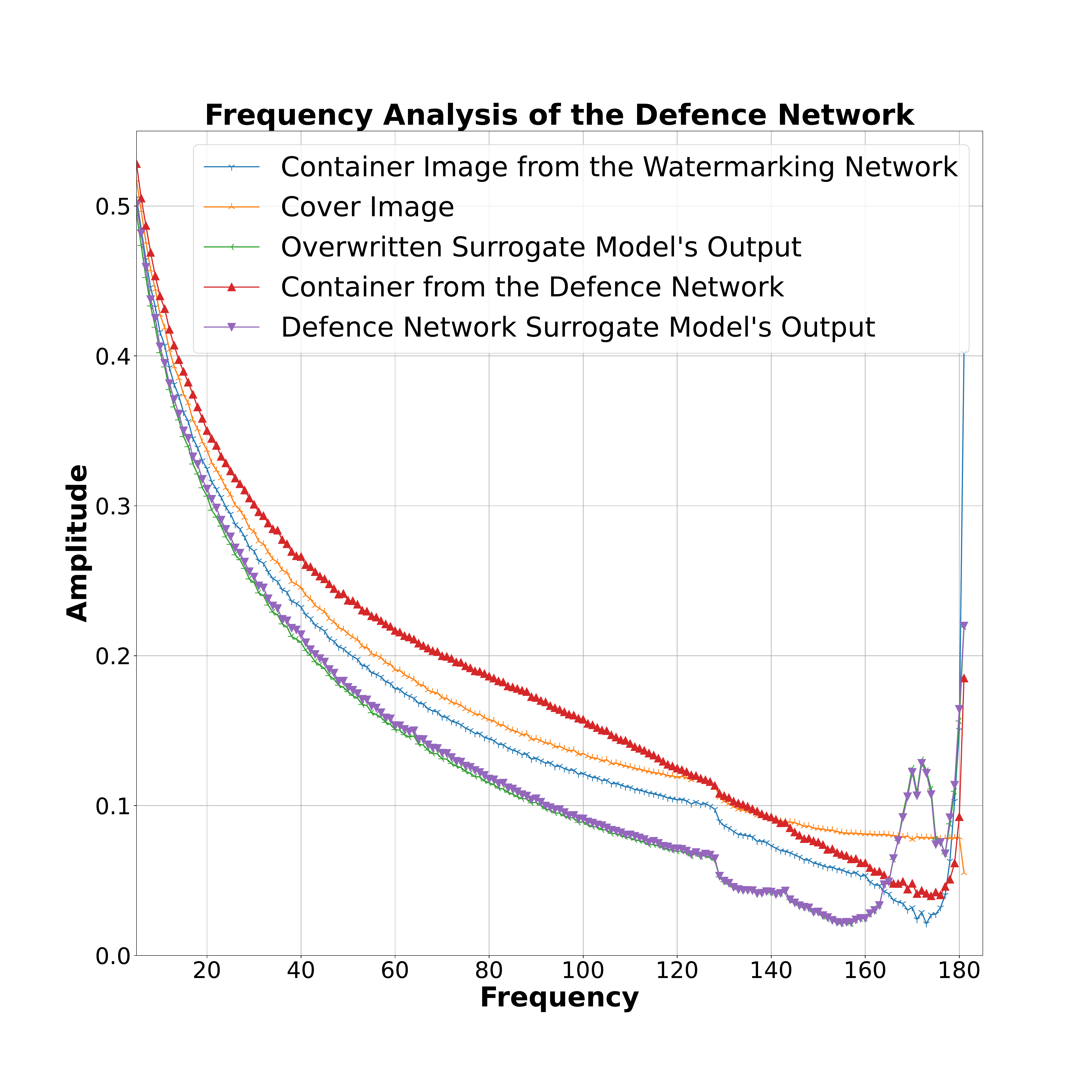}
        \caption{The defense Method}
        \label{fig:freq_als_def}
    \end{subfigure}
    
    \caption{Frequency Analysis}
    \label{fig:freq_als}
\end{figure*}





\subsubsection{Frequency Analysis}

The objective of the study is to investigate the cause of the overwriting attack's capability to render the embedded watermark in the container image ineffective. This is achieved by calculating the Azimuthal Integral of the experimental images and comparing their frequency domains. The final data is obtained by averaging the Azimuthal Integral computed from 1,000 groups of test images, each group consisting of the container image generated by the watermarking network, the cover image, the overwritten container image, the output from the surrogate model, and the overwritten output from the surrogate model. The images within each group correspond to the same processed image.

Typically, images processed by Deep Convolutional Neural Networks (DCNNs) display a bias in the high frequency domain. As illustrated in Figure \ref{fig:freq_als_atk}, the container image generated by the watermarking network and its corresponding image generated by the surrogate model exhibit an abnormally high amplitude in the high frequency domain, which distinguishes them greatly from the cover image. This is the reason why the watermark can be invisibly embedded into the cover image, as human eyes are not sensitive enough to the high frequency domain of an image.

However, through fine-tuning, the retrieval network in the watermarking network can still retrieve the watermark from the surrogate model's output, despite its significant deviation from the frequency distribution of the container image. This emphasizes the significance of the fine-tuning stage. In the case of the overwritten container image, it displays a marked bias in the high frequency domain, both in comparison to the cover image and the watermarked image. A peak can be observed in the range of 160 to 175 on the frequency axis, which neutralizes the previously embedded watermark.

\begin{figure}[t!]
    \centering
    \includegraphics[width=.3\textwidth]{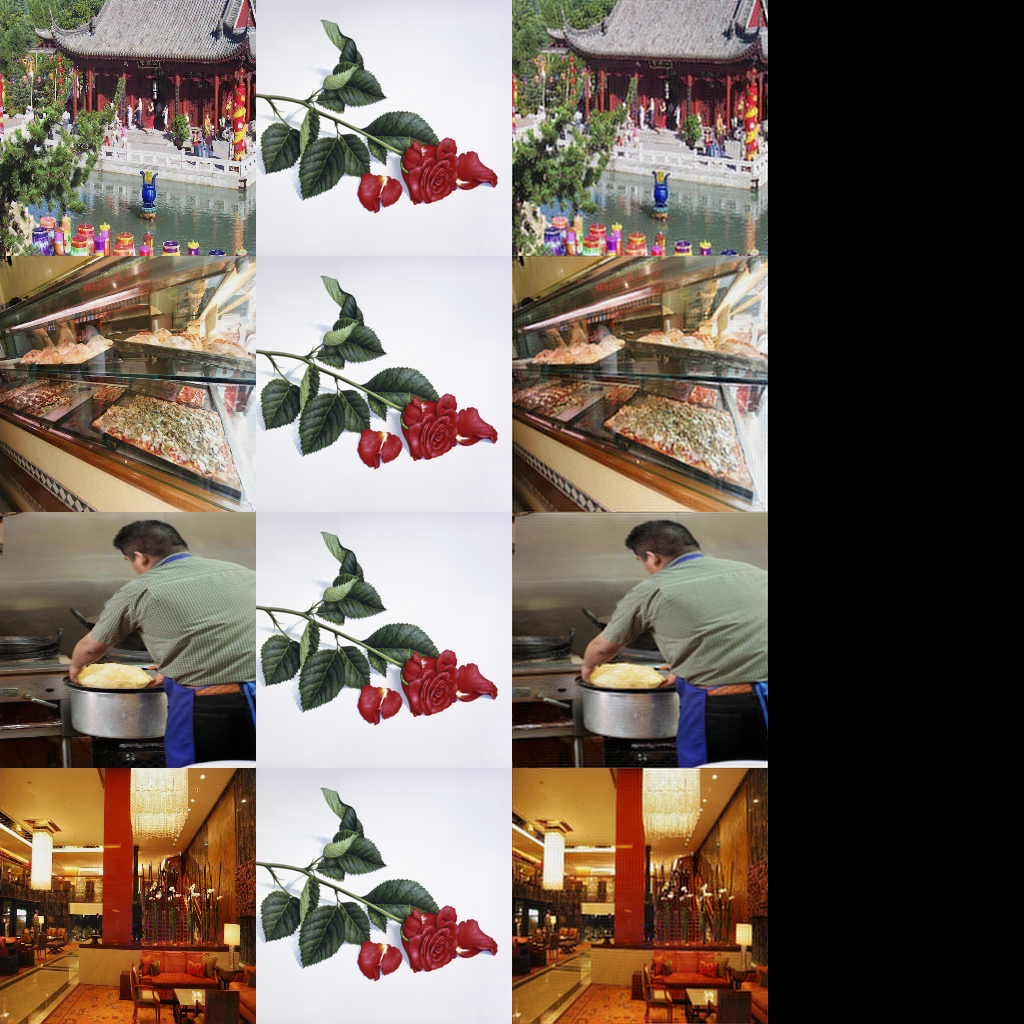}
    \caption{Watermark Retrieval of Low-pass Filtered Container Images}
    \label{fig:low_pass}
\end{figure}

To further ascertain the location where the watermark is embedded, a low-pass filter is applied to the watermarked images. The filtered image retains its visual quality to the extent that changes are not easily noticeable by the human eye. This filter is applied to 1,000 container images and then the watermark retrieval is performed. As expected, the success rate of the retrieval drops to 0, and the direct effect can be seen in Figure \ref{fig:low_pass}, where each column in each row, from left to right, respectively represents the container image, the retrieved watermark, the filtered container image, and the nullified retrieval result. This underscores the high sensitivity of the watermark retrieval to the high frequency distribution of the container image.


\subsection{Analysis of the defense Network}





\subsubsection{Residue Analysis}
\begin{figure}[t!]
    \centering
    
    \begin{subfigure}[b]{0.24\textwidth}
        \centering
        \includegraphics[width=\textwidth]{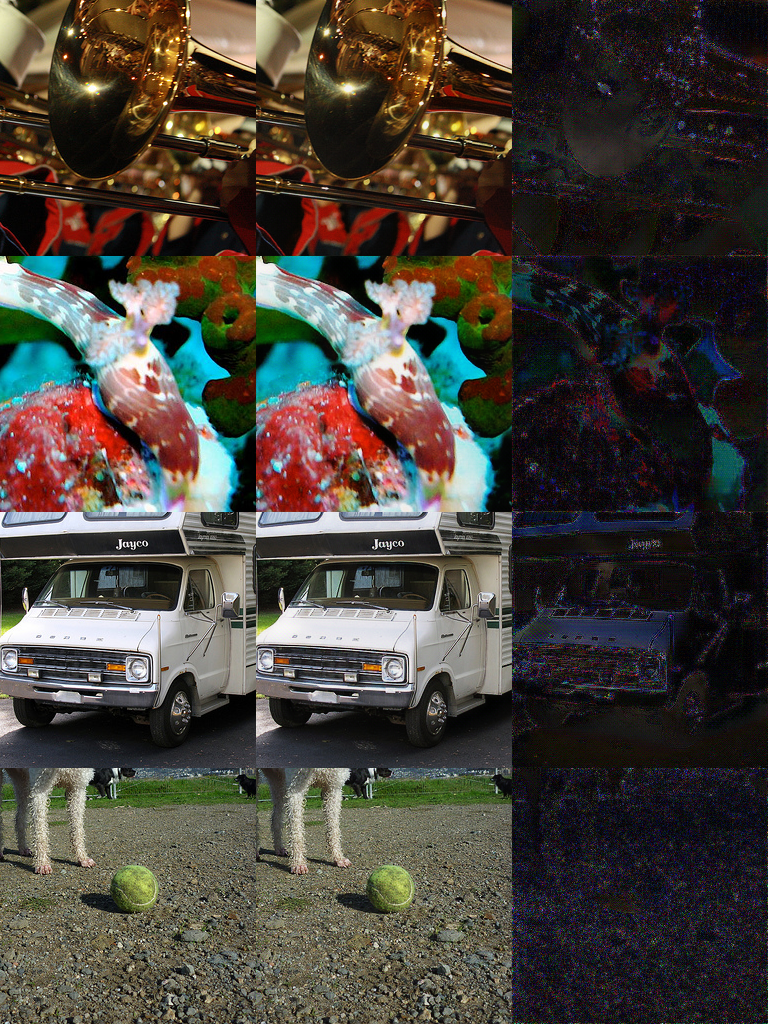}
        \caption{defense Network}
    \end{subfigure}
    \hfill
    \begin{subfigure}[b]{0.24\textwidth}
        \centering
        \includegraphics[width=\textwidth]{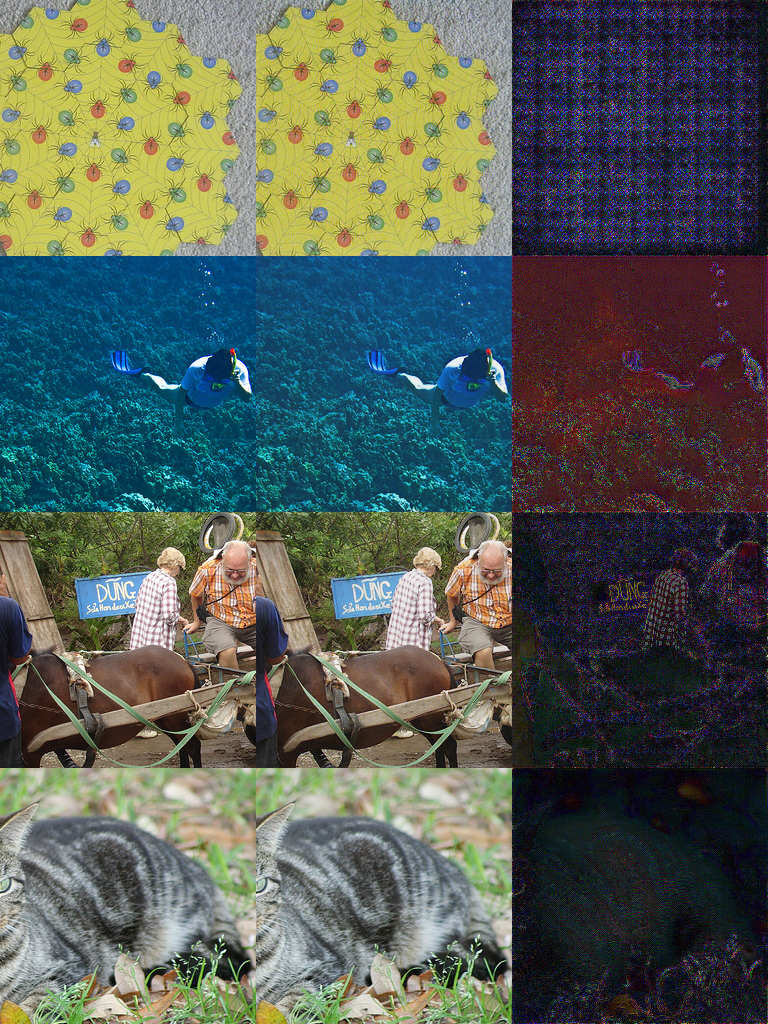}
        \caption{Watermarking Network}
    \end{subfigure}
    
    \caption{Residue Analysis (10x Enhanced)}
    \label{fig:residue_als}
\end{figure}

First, we performed a residue analysis on the container images generated by our defense and watermarking networks.
The details can be seen in Figure \ref{fig:residue_als}, where from left to right, each column in each row respectively represents the cover image, the container image, and the residue enhanced 10 times.
Intuitively, the residues of the defense network's output seem to be darker (better) than those of the watermarking network's output.

\subsubsection{Frequency Analysis.}
In the adversarial stage of the watermarking network, the retrieval network is required to retrieve the watermark from both the surrogate model's output and the container images.
Due to the bias in the frequency domain of the overwritten surrogate model's output shown in Figure \ref{fig:freq_als_atk}, the retrieval fails, because it has never encountered any input with such a frequency distribution.
However, in the defense network's fine-tuning stage, the surrogate model's output is protected by the defense network, and they share almost the same frequency distribution as the overwritten surrogate model's output in Figure \ref{fig:freq_als_def}.
This forces the retrieval network to become more robust to the mutable watermark signal.
Further, with the assistance of the frequency loss, the container images generated by the defense network share a more similar frequency distribution to the cover images than those generated by the watermarking network.
Our defense method therefore shows good robust to overwriting attack, even if the type of the surrogate model does not match the one used in fine-tuning.
Nevertheless, it is still possible for an adversary to change the attack method to cripple the watermark with a higher cost of visual quality – through low-pass filtering for example.

\subsection{Ablation Study}


    
    

\subsubsection{The frequency loss}
The ablation experiments in Zhang {\it et al.} \cite{zhang2020model} prove the necessity of several loss terms, including the clean loss and the consistent loss.
In our defense network, the frequency loss regularizer is added into the loss function so as to guide the network to generate the container images that share a more similar frequency distribution to the cover image.
The only difference between the loss terms in our defense network and the watermarking network is the frequency loss.
This boosts the image quality, as is proven in the test results presented in Tables \ref{tab:res_org}, and \ref{tab:res_def}. 
Here, both the PSNR and the SSIM values of the container images generated by our defense network are higher than those from the watermarking network.
Further, as Figure \ref{fig:freq_als_def} shows, the high frequency distribution of the containers from the defense network is closer to the cover image than those from the watermarking network.

\subsubsection{Fine-tuning}
Unlike the original watermarking method in \cite{zhang2020model}, we additionally add a group of watermark-free images generated by the surrogate model trained on the watermark-free training dataset into the fine-tuning dataset.
This prevents the watermarking network from overfitting the steganographic signal in the container images so that it will retrieve the watermark regardless of what exact watermark signal lies in the container images.
Figure \ref{fig:overfitting} shows how the overfitting phenomenon nullifies the watermarking method.
Therefore, the watermark-free surrogate model's output is essential in this stage.
If a watermark can be retrieved from a container image that does not contain the specific watermarking signal, the method can be claimed unreliable.
Further, by inserting the overwriting network into the watermarking network to form the defense network, the defense network is pushed to become more robust to both the overwriting attack and the addition of noise.
Further, the embedding network hides the watermark more covertly, and the retrieval network differentiates the container image carrying both the specific watermark and the overwritten watermark images from the watermark-free images, and the watermark-free surrogate model's output, and the container images carrying any other secret images.



\section{Conclusion}
\label{sect:con}



In this study, we present an overwriting attack that effectively nullifies the watermark embedded in images processed by image processing neural networks. Our attack is also a threat to deep steganography, as it can invisibly replace a secret image with minimal impact on the visual quality of the image. Additionally, we identify an overfitting issue in the original watermarking method and resolve it with an alternative training approach.
To defend against our proposed overwriting attack, we develop an adversarial framework defense network that integrates the watermarking network and the overwriting network. To the best of our knowledge, this defense network is resilient against the overwriting attack. Through adversarial training, the defense network is able to retrieve valid watermarks from overwritten images and the output of the overwritten surrogate model.

There is ample room for future research in the area of image-processing model watermarking, including the development of robust watermarking techniques and malicious attacks. Although our method demonstrates robustness against overwriting attacks, the adversary can still manipulate the frequency domain of the output to erase the embedded watermark with minimal perceptual impact. To address this issue, a more robust watermarking method that embeds the watermark in the low frequency domain of the image should be explored.

\bibliographystyle{IEEEtran}
\bibliography{IEEEabrv,ref}
\end{document}